\pdfoutput=1
\RequirePackage{ifpdf}
\ifpdf 
\documentclass[pdftex]{sigma}
\else
\documentclass{sigma}
\fi

\DeclareMathAlphabet{\mymathbb}{U}{BOONDOX-ds}{m}{n}
\numberwithin{equation}{section}

\newtheorem*{Theorem*}{Theorem}

 { \theoremstyle{definition}

 }

\newcommand\Aut{\operatorname{Aut}}
\newcommand\im{\operatorname{Im}}
\newcommand\Ker{\operatorname{Ker}}

\begin{document}
\allowdisplaybreaks

\newcommand{\arXivNumber}{2206.11138}

\renewcommand{\PaperNumber}{024}

\FirstPageHeading

\ShortArticleName{Some Useful Operators on Differential Forms on Galilean and Carrollian Spacetimes}

\ArticleName{Some Useful Operators on Differential Forms\\ on Galilean and Carrollian Spacetimes}

\Author{Mari\'an FECKO}

\AuthorNameForHeading{M.~Fecko}

\Address{Department of Theoretical Physics, Comenius University in Bratislava, Slovakia}
\Email{\href{mailto:Marian.Fecko@fmph.uniba.sk}{Marian.Fecko@fmph.uniba.sk}}

\ArticleDates{Received August 30, 2022, in final form April 11, 2023; Published online April 22, 2023}

\Abstract{Differential forms on Lorentzian spacetimes is a well-established subject. On Galilean and Carrollian spacetimes it does not seem to be quite so. This may be due to the absence of Hodge star operator. There are, however, potentially useful analogs of Hodge star operator also on the last two spacetimes, namely intertwining operators between corresponding representations on forms. Their use could perhaps make differential forms as attractive tool for physics on Galilean and Carrollian spacetimes as forms on Lorentzian spacetimes definitely proved to be.}


\Classification{53Z05; 83C99; 22E70}

\section{Introduction and motivation}\label{intro}

A lot of important mathematics routinely used in physics reduces to manipulations with \emph{differential forms}
on (pseudo)-\emph{Riemannian} manifolds
(see, e.g., \cite{BambergSternberg, CrampinPirani, Feckoangl, Frankel123l, GockSchuck, Schutz, Trautman}).

In particular, on 3-dimensional Euclidean space, $E^3$, the whole standard vector calculus
more or less boils down to simple manipulations with combinations of just two operators on forms,
the exterior derivative $d$ and the \emph{Hodge star} operator $*$.

Similarly on Minkowski spacetime $E^{1,3}$, expression of, say, Maxwell equations
and its consistency condition (local conservation of electric charge),
\[
 {\rm d}*F = - J,
 \qquad
 {\rm d}F = 0 ,
 \qquad
 {\rm d}J = 0 ,
\]
the wave operator $(\sim *{\rm d}*{\rm d})$,
Lorentz gauge condition ($*{\rm d}*A=0$) etc., again only use forms and the above mentioned two operators, ${\rm d}$ and $*$.

A lot of research is recently done on \emph{Galilean} and \emph{Carrollian} spacetimes,
see, e.g., \cite{Banerjeeetal2020, Ciambellietal2019, DuvGibHorZha2014, Figueroa2020,Figueroa2019, HenneaxSalReb2021, Morand2018, BleekenYunus2016} or
a long lists of references in recent Ph.D.~Theses~\cite{GrassiePhD} and~\cite{HansenPhD}.
What about using forms (together with the two operators, ${\rm d}$ and $*$), there?
It turns out, it cannot be done in a~straightforward way.

Let us illustrate it on the simplest versions of the two spacetimes, \emph{Galilei} and \emph{Carroll} spacetimes.
(Galilean and Carrollian spacetimes are \emph{locally} Galilei and Carroll, respectively,
 in a~simi\-lar way as usual curved spacetimes, Lorentzian manifolds, are \emph{locally} Lorentz.
 See Appendices~\ref{galileicarrollvector} and~\ref{galileancarrollianmanifold}.)
We can treat them as just $\mathbb R^4$ with global Cartesian coordinates $(t,x,y,z) \equiv (t,\mathbf r)$,
where translations and rotations act exactly as in Minkowski spacetime, but boosts act differently, namely
\begin{alignat}{5} 
 \label{tgalcarflat}
 &\text{Galilei boost}\colon \ && t' = t,\qquad &&
 \text{Carroll boost}\colon \ && t' = t + \mathbf v \cdot \mathbf r,& \\
 \label{rgalcarflat}
&&& \mathbf r' = \mathbf r + \mathbf v t ,\qquad &&&&
 \mathbf r' = \mathbf r&
\end{alignat}
(Galilei and Carroll spacetimes may be treated as $c\to \infty$ and $c\to 0$ limits
 of the Minkowski spacetime, respectively.)

Now one can easily check that there is neither Galilei-invariant \emph{metric tensor} on Galilei spacetime
nor Carroll-invariant metric tensor on Carroll spacetime
 (analogs of Minkowski metric tensor on Minkowski spacetime).\footnote{So Hermann Minkowski could not construct useful spacetime metric prior to the change of paradigm
 from Galilei relativity to Einstein relativity.}

Recall, however, that the Hodge star operator \emph{needs metric tensor} (and orientation) for its definition
\[ 
 *\colon \ \Omega^p \to \Omega^{n-p},
 \qquad
 \Omega^p \equiv \Omega^p(M,g,o)
\] 
and it behaves according to
\[ 
 f^* *_{g,o} = *_{f^*g,f(o)} f^*
\] 
w.r.t.\ diffeomorphisms; it therefore \emph{commutes} with $f^*$ for (orientation preserving) isometries
\[ 
 f^*g = g
 \Rightarrow
 f^* *_g = *_g f^*.
\] 
In particular it is translation and rotation-invariant (in this sense) on $E^3$ and Poincar\'e-invariant on Minkowski space.

The lack of Galilei/Carroll invariant metric tensors on Galilei/Carroll spacetimes makes it impossible to construct
Galilei/Carroll invariant Hodge star operator in \emph{standard} way.

We can, however, try to find useful \emph{analogs} of the Hodge star. Namely to find \emph{directly} op\-er\-a\-tors
\[ 
 \text{``}*\text{''}\colon\ \Omega^p \to \Omega^q,
\qquad
 \Omega^p \equiv \Omega^p \quad (\text{Galilei/Carroll})
\] 
(we are ready to find such operators for \emph{any pair} of degrees rather than just $p$ and $(n-p)$)
acting on forms on Galilei/Carroll spacetime so that
\begin{equation} \label{Hodgegalcarproperty}
 f^* \text{``}*\text{''} = \text{``}*\text{''} f^*
\qquad \text{whenever $f$ is Galilei/Carroll transformation}
\end{equation}
(i.e., boost from (\ref{tgalcarflat})--(\ref{rgalcarflat}) or rotation).

Corresponding problem is addressed in Sections \ref{FormsGalCarMink}--\ref{aqpminkgalcarnontrivial}.
Its solution is based on observation, that Galilei/Car\-roll transformations on forms
induce \emph{representations} of Galilei/Carroll groups (and, consequently, of the Lie algebras)
on the \emph{space of components} of forms $\hat \Lambda^p$
and the condition~(\ref{Hodgegalcarproperty}) is reflected in validity of the commutative diagram
\begin{equation}
\begin{CD} \label{intertwinptoqintro}
 \hat \Lambda^p @>\hat a_{qp}>> \hat \Lambda^q \\
 @V{\rho_p}VV @VV{\rho_q}V \\
 \hat \Lambda^p @>>\hat a_{qp}> \hat \Lambda^q \\
\end{CD}
\qquad \text{i.e.,} \quad
 \rho_q \circ \hat a_{qp} = \hat a_{qp} \circ \rho_p.
\end{equation}
So we actually look for \emph{intertwining operators} for some finite-dimensional representations.

In what follows we will see that such operators do exist and that their list depends heavily on the choice of particular spacetime
(Galilei, Carroll, Minkowski).
In particular, in Minkowski spacetime, the \emph{only} (non-trivial) intertwining ope\-ra\-tor
turns out to be the Hodge star.
In the Galilei and Carroll spacetimes, on the contrary, there are operators
\begin{enumerate}\itemsep=0pt\samepage
\item[--] acting similarly to (Minkowski) Hodge star (already known from \cite{FeckoGalCarHodge}),
\item[--] with no similarity to (Minkowski) Hodge star.
\end{enumerate}
So, in this sense, the structure of intertwining operators on forms is slightly \emph{richer} in
the Galilei and Carroll spacetimes than it is in the Minkowski one.

Now the \emph{standard} Hodge star $*_{g,o}$ on a general $(M,g,o)$ happens to be actually invariant even
w.r.t.\ \emph{local} isometries
(it is \emph{local Lorentz} invariant on a general-relativistic spacetime).
So its importance goes far beyond isometries of $(M,g,o)$.

Therefore, from the perspective of usefulness in \emph{Galilean and Carrollian} spacetimes,
we look for \emph{analogs} of Hodge star which enjoy the property of \emph{local} Galilei/Carroll invariance.
This can be, however, already achieved (from results obtained in Sections \ref{FormsGalCarMink}--\ref{aqpminkgalcarnontrivial})
by standard techniques ($G$-structures, see
Appendices \ref{galileicarrollvector}, \ref{galileancarrollianmanifold} and \ref{spacetimevectorspace}).

So the results obtained for Galilei/Carroll cases remain valid in \emph{Galilean and Carrollian} spacetimes as well.

\subsection{Structure of the paper}
\label{structure}

In Section \ref{FormsGalCarMink}, we first introduce standard parametrization of differential forms
in terms of their~``1+3'' decompositions.
It is common for Minkowski, Galilei and Carroll spacetimes.
In this language, generators $\rho_p = (S_j,N_j)$ of the representations of the corresponding three Lie algebras
on component spaces of $p$-forms are explicitly computed. Boosts generators $N_j$ depend on particular spacetime.

Using the matrices of the generators, explicit matrix equations for the intertwining operators~$a_{qp}$ are written
in Section \ref{intertwiningaqp}. They are simplified a lot due to rotations.
Typically only a~few free constants survive. The constants are then further restricted by boosts.

Resulting nontrivial intertwining matrices are presented in Section \ref{aqpminkgalcarnontrivial}.
They are also translated back into the original language of ``1+3'' decomposed differential forms
(as is convenient to see them in physics).

Section \ref{summaryconclusions} provides summary and conclusions.

Appendix \ref{detailssolving} presents detailed calculation of (more or less random) subset of matrices $a_{qp}$.

In Appendix \ref{app:galcarminhodgefromFecko}, list of action of \emph{Hodge star} operators on forms
is given. They were computed in \cite{FeckoGalCarHodge} by completely different method
(appropriate \emph{modification} of standard Hodge star \emph{formula}, using specific invariant tensors
 present on Galilean and Carrollian spacetimes). They are useful for comparison of the results.

Appendices \ref{galileicarrollvector} and
 \ref{galileancarrollianmanifold}
collect, for convenience of the reader, all information on how we proceed,
starting from Galilei and Carroll vector spaces, to Galilean and Carrollian structure on manifolds
(i.e., how to make each \emph{tangent} space of $M$ a \emph{Galilei and Carroll} vector space).
And how intertwining operators are connected for the two situations.

Appendix \ref{spacetimevectorspace} discusses in more detail why computations performed
in Sections \ref{FormsGalCarMink}--\ref{aqpminkgalcarnontrivial} (concerning Galilei and Carroll spacetimes)
may be more or less directly used in order to formulate corresponding statements
for forms on Galilean and Carrollian spacetimes.

Finally Appendix \ref{spacetimeHodge} provides explicit expressions for the $*$ operator
on general Galilean and Carrollian spacetimes.

\section{Forms on Galilei, Carroll and Minkowski spacetime}
\label{FormsGalCarMink}

\subsection{Parametrization}
\label{param}

In order to write down explicitly equations (\ref{intertwinptoqintro}),
we need to determine explicit form of the matrices $\rho_p$ for
\begin{enumerate}\itemsep=0pt
\item[--] all three $(1+3)$-dimensional spacetimes (Galilei, Carroll, Minkowski),
\item[--]all degrees $p=0,1,2,3,4$ of differential forms.
\end{enumerate}
Let us start with a suitable coordinates and parametrization of forms.

In all three cases we have (global) coordinates $\big(t,x^i\big) = (t,x,y,z)$ adapted to particular spacetime structures.
In these coordinates, we can use, in \emph{all three} cases,
parametrization of differential forms well-known from the Minkowski spacetime
(see, e.g., \cite[Section 16.1]{Feckoangl}). That is, any $p$-form $\alpha$,
$p=0,1,2,3,4$, may be first uniquely decomposed as
\begin{equation} \label{alphadecomp}
 \alpha = {\rm d}t \wedge \hat s + \hat r,
\end{equation}
where the two hatted forms, $(p-1)$-form $\hat s$ and $p$-form $\hat r$, respectively, are \emph{spatial},
meaning that they do not contain (``temporal'') differential ${\rm d}t$ in its coordinate expression.
Being spatial, $\hat s$~and~$\hat r$ may be, depending on degree, expressed as
\[ 
 f, \qquad
 \mathbf a \cdot {\rm d}\mathbf r, \qquad
 \mathbf a \cdot {\rm d}\mathbf S, \qquad
 f{\rm d}V,
\] 
where
\begin{gather*} 
 \mathbf a \cdot {\rm d}\mathbf r
 = a_x{\rm d}x + a_y{\rm d}y + a_z{\rm d}z,\\
 \mathbf a \cdot {\rm d}\mathbf S
 = a_x{\rm d}S_x + a_y{\rm d}S_y + a_z{\rm d}S_z
 \equiv a_x{\rm d}y\wedge {\rm d}z + a_y{\rm d}z \wedge {\rm d}x + a_z{\rm d}x \wedge {\rm d}y, \\
 {\rm d}V
 = {\rm d}x \wedge {\rm d}y \wedge {\rm d}z.
\end{gather*}
So, the general decomposition formula (\ref{alphadecomp}) may be further specified,
for the five relevant degrees of differential forms, as follows:
\begin{gather} 
 \label{Omega0}
 \Omega^0 \colon \ \alpha
 = f, \\
 \label{Omega1}
 \Omega^1 \colon \ \alpha
= f{\rm d}t + \mathbf a \cdot {\rm d}\mathbf r, \\
 \label{Omega2}
 \Omega^2 \colon \ \alpha
 = {\rm d}t \wedge \mathbf a \cdot {\rm d}\mathbf r + \mathbf b \cdot {\rm d}\mathbf S, \\
 \label{Omega3}
 \Omega^3 \colon \ \alpha
 = {\rm d}t \wedge \mathbf a \cdot {\rm d}\mathbf S + f{\rm d}V, \\
 \label{Omega4}
 \Omega^4 \colon \ \alpha
 = f{\rm d}t \wedge {\rm d}V
\end{gather}
(Notice that the time coordinate $t$ itself may be present in \emph{components} of the spatial parts,
 for example $a_x = a_x(t,x,y,z)$ in general.)

Alternatively, we can uniquely identify forms from (\ref{Omega0})--(\ref{Omega4}) just with columns
\begin{gather} \label{columnsformsgalcarlor}
 f,
\qquad
 \left( \begin{matrix}
 f \\
 \mathbf a
 \end{matrix}
 \right),
\qquad
 \left( \begin{matrix}
 \mathbf a \\
 \mathbf b
 \end{matrix}
 \right),
\qquad
 \left( \begin{matrix}
 f \\
 \mathbf a
 \end{matrix}
 \right),
\qquad
 f.
\end{gather}
Now we go on to study action of corresponding spacetime symmetry groups on forms within this parametrization.

\subsection{Generators of rotations}
\label{generrot}
In the above mentioned coordinates, what is common for all three spacetimes is action of
\begin{enumerate}\itemsep=0pt
\item[--] time and space translation,
\item[--]spatial (3D) rotations
\begin{gather} 
 \label{translrotcommont}
 t' = t + t_0, \\
 \label{translrotcommonr}
 \mathbf r' = R\mathbf r + \mathbf r_0,
\end{gather}
\end{enumerate}
where $R$ denotes $3\times 3$ rotation matrix.

What is different, is action of boosts (see below, Section \ref{generboost}).

Concerning scrambling of components of forms, time and space \emph{translations} alone
do \emph{nothing} (since ${\rm d}t_0$ and ${\rm d}\mathbf r_0$ vanish), so we can forget about them
and restrict to rotations and boosts.

Moreover, we can restrict to \emph{infinitesimal} rotations and boosts since
even if $\rho_p$ in the commutative diagram (\ref{intertwinptoqintro}) were meant
as operators representing corresponding symmetry \emph{Lie group},
the same intertwining operators enter the ``infinitesimal version''
of the commutative diagram, where $\rho_p$ already denote operators of the
\emph{derived} representation of the corresponding symmetry \emph{Lie algebra}.

For infinitesimal rotations about $j$-th axis,
formulas (\ref{translrotcommont})--(\ref{translrotcommonr}) reduce to
\begin{equation} \label{infrotjmatrixform}
 \left( \begin{matrix}
 t' \\
 \mathbf r'
 \end{matrix}
 \right)
 =
 \left( \begin{matrix}
 t \\
 \mathbf r
 \end{matrix}
 \right)
 + \epsilon
 \left( \begin{matrix}
 0 & 0 \\
 0 & l_j
 \end{matrix}
 \right)
 \left( \begin{matrix}
 t \\
 \mathbf r
 \end{matrix}
 \right),
\end{equation}
where
\[ 
 (l_j)_{ik} = \epsilon_{ijk}.
\] 
Formula (\ref{infrotjmatrixform}) induces action on coordinate basis of forms
\begin{gather} 
 \label{rotdt}
 {\rm d}t' = {\rm d}t, \\
 \label{rotdr}
 {\rm d}\mathbf r' = (\mymathbb{1} + \epsilon l_j){\rm d}\mathbf r, \\
 \label{boostlorinfdtdr}
 {\rm d}\mathbf S' = (\mymathbb{1} + \epsilon l_j){\rm d}\mathbf S, \\
 \label{boostlorinfdS}
 {\rm d}V' = {\rm d}V.
\end{gather}
From requirement
\begin{gather} 
 \label{rotOmega0prime}
 f' = f, \\
 \label{rotOmega1prime}
 f'{\rm d}t' + \mathbf a' \cdot {\rm d}\mathbf r' = f{\rm d}t + \mathbf a \cdot {\rm d}\mathbf r, \\
 \label{rotOmega2prime}
 {\rm d}t' \wedge \mathbf a' \cdot {\rm d}\mathbf r' + \mathbf b' \cdot {\rm d}\mathbf S' = {\rm d}t \wedge \mathbf a \cdot {\rm d}\mathbf r + \mathbf b \cdot {\rm d}\mathbf S, \\
 \label{rotOmega3prime}
 {\rm d}t' \wedge \mathbf a' \cdot {\rm d}\mathbf S' + f'{\rm d}V' = {\rm d}t \wedge \mathbf a \cdot {\rm d}\mathbf S + f{\rm d}V, \\
 \label{rotOmega4prime}
 f'{\rm d}t' \wedge {\rm d}V' = f{\rm d}t \wedge {\rm d}V,
\end{gather}
we can compute, for each degree, primed components in terms of unprimed and get
\begin{align}
 \label{rot0comp}
 &\Omega^0\colon \ f'
 =
 f, \\
 \label{rot1comp}
 &\Omega^1\colon \ \left( \begin{matrix}
 f' \\
 \mathbf a'
 \end{matrix}
 \right)
 =
 \left( \begin{matrix}
 f \\
 \mathbf a
 \end{matrix}
 \right)
 + \epsilon
 \left( \begin{matrix}
 0 & 0 \\
 0 & l_j
 \end{matrix}
 \right)
 \left( \begin{matrix}
 f \\
 \mathbf a
 \end{matrix}
 \right), \\
 \label{rot2comp}
&\Omega^2\colon \
 \left( \begin{matrix}
 \mathbf a' \\
 \mathbf b'
 \end{matrix}
 \right)
 =
 \left( \begin{matrix}
 \mathbf a \\
 \mathbf b
 \end{matrix}
 \right)
 + \epsilon
 \left( \begin{matrix}
 l_j & 0 \\
 0 & l_j
 \end{matrix}
 \right)
 \left( \begin{matrix}
 \mathbf a \\
 \mathbf b
 \end{matrix}
 \right), \\
 \label{rot3comp}
 &\Omega^3\colon \
 \left( \begin{matrix}
 f' \\
 \mathbf a'
 \end{matrix}
 \right)
 =
 \left( \begin{matrix}
 f \\
 \mathbf a
 \end{matrix}
 \right)
 + \epsilon
 \left( \begin{matrix}
 0 & 0 \\
 0 & l_j
 \end{matrix}
 \right)
 \left( \begin{matrix}
 f \\
 \mathbf a
 \end{matrix}
 \right), \\
 \label{rot4comp}
 &\Omega^4\colon \
 f'
 =
 f.
\end{align}
All the formulas may be written (for rotation about $j$-th axis)
as multiplication of the correspondent component column by matrix
\[ 
\mymathbb{1} + \epsilon S_j,
\] 
where
\begin{align*} 
 &\Omega^0\colon \ S_j =
 0, \\
 &\Omega^1\colon \ S_j =
 \left( \begin{matrix}
 0 & 0 \\
 0 & l_j
 \end{matrix}
 \right), \\
&\Omega^2\colon \ S_j =
 \left( \begin{matrix}
 l_j & 0 \\
 0 & l_j
 \end{matrix}
 \right), \\
 &\Omega^3\colon \ S_j =
 \left( \begin{matrix}
 0 & 0 \\
 0 & l_j
 \end{matrix}
 \right), \\
 &\Omega^4\colon \ S_j = 0.
\end{align*}

\subsection{Generators of boosts}
\label{generboost}

Let us display all three kinds of infinitesimal boosts along $\mathbf n$:
\begin{alignat} {3}
 \label{boostlort}
 &\text{Lorentz}\colon \ &&
 t' = t + \epsilon \mathbf n \cdot \mathbf r,& \\
 \label{boostlorr}
&&&\mathbf r' = \mathbf r + \epsilon \mathbf n t,&
\\
 \label{boostgalt}
 &\text{Galilei} \colon \ &&
 t' = t,& \\
 \label{boostgalr}
&&&\mathbf r' = \mathbf r + \epsilon \mathbf n t,&
\\
 &\text{Carroll} \colon \ &&
 \label{boostcarinft}
 t' = t + \epsilon \mathbf n \cdot \mathbf r,& \\
 \label{boostcarinfr}
&&&\mathbf r' = \mathbf r.&
\end{alignat}
(As for the Carroll boosts, see \cite[formula (9)]{LevyLeblond}, \cite[formula~(2.10)]{DuvGibHorZha2014}
 or Appendix \ref{carrollvectorbasics}.)
The following formulas conveniently cover all three cases at once:
\begin{gather*} 
 t' = t + \epsilon \lambda_1 \mathbf n \cdot \mathbf r, \qquad
 \mathbf r' = \mathbf r + \epsilon \lambda_2 \mathbf n t.
\end{gather*}
The three special cases clearly correspond to values
\begin{alignat}{3} 
 \label{lorlambdas}
& \text{Lorentz}\colon \ &&
 (\lambda_1,\lambda_2) = (1,1), &\\
 \label{gallambdas}
 &\text{Galilei}\colon \ &&
 (\lambda_1,\lambda_2) = (0,1),& \\
 \label{carlambdas}
& \text{Carroll}\colon \ &&
 (\lambda_1,\lambda_2) = (1,0).&
\end{alignat}
Induced action on coordinate basis of forms comes out to be
\begin{alignat}{3} 
 \label{lorgalcarboostinfdt}
 &\Omega^1 \colon \ && {\rm d}t' = {\rm d}t + \epsilon \lambda_1 \mathbf n \cdot {\rm d}\mathbf r,& \\
 \label{lorgalcarboostinfdr}
 &&& {\rm d}\mathbf r' = {\rm d}\mathbf r + \epsilon \lambda_2 \mathbf n {\rm d}t;& \\
 \label{lorgalcarboostinfdtdr}
 &\Omega^2 \colon \ && {\rm d}t' \wedge {\rm d}\mathbf r' = {\rm d}t \wedge {\rm d}\mathbf r - \epsilon \lambda_1 (\mathbf n \cdot \mathbf l){\rm d}\mathbf S,& \\
 \label{lorgalcarboostinfdS}
 &&& {\rm d}\mathbf S' = {\rm d}\mathbf S + \epsilon \lambda_2 {\rm d}t \wedge (\mathbf n \cdot \mathbf l){\rm d}\mathbf r;& \\
 \label{lorgalcarboostinfdtdS}
 &\Omega^3 \colon \ && {\rm d}t' \wedge {\rm d}\mathbf S' = {\rm d}t \wedge {\rm d}\mathbf S + \epsilon \lambda_1 \mathbf n {\rm d}V,& \\
 \label{lorgalcarboostinfdV}
 &&& {\rm d}V' = {\rm d}V + \epsilon \lambda_2 {\rm d}t \wedge (\mathbf n \cdot {\rm d}\mathbf S);& \\
 \label{lorgalcarboostinfdtdV}
 &\Omega^4 \colon \ && {\rm d}t' \wedge {\rm d}V' = {\rm d}t \wedge {\rm d}V.&
\end{alignat}
Writing down again (\ref{rotOmega0prime})--(\ref{rotOmega4prime}) we get boost (along $x_j$) analogs of
(\ref{rot0comp})--(\ref{rot4comp})
\begin{align} 
 \label{boost0comp}
 &\Omega^0\colon \ f'
 =
 f, \\
 \label{boost1comp}
 &\Omega^1\colon \ \left( \begin{matrix}
 f' \\
 \mathbf a'
 \end{matrix}
 \right)
 =
 \left( \begin{matrix}
 f \\
 \mathbf a
 \end{matrix}
 \right)
 + \epsilon
 \left( \begin{matrix}
 0 & \lambda_2 e_j^{\rm T} \\
 \lambda_1 e_j & 0
 \end{matrix}
 \right)
 \left( \begin{matrix}
 f \\
 \mathbf a
 \end{matrix}
 \right), \\
 \label{boost2comp}
&\Omega^2\colon \
 \left( \begin{matrix}
 \mathbf a' \\
 \mathbf b'
 \end{matrix}
 \right)
 =
 \left( \begin{matrix}
 \mathbf a \\
 \mathbf b
 \end{matrix}
 \right)
 + \epsilon
 \left( \begin{matrix}
 0 & - \lambda_2 l_j \\
 \lambda_1 l_j & 0
 \end{matrix}
 \right)
 \left( \begin{matrix}
 \mathbf a \\
 \mathbf b
 \end{matrix}
 \right), \\
 \label{boost3comp}
 &\Omega^3\colon \
 \left( \begin{matrix}
 f' \\
 \mathbf a'
 \end{matrix}
 \right)
 =
 \left( \begin{matrix}
 f \\
 \mathbf a
 \end{matrix}
 \right)
 + \epsilon
 \left( \begin{matrix}
 0 & \lambda_1 e_j^{\rm T} \\
 \lambda_2 e_j & 0
 \end{matrix}
 \right)
 \left( \begin{matrix}
 f \\
 \mathbf a
 \end{matrix}
 \right), \\
 \label{boost4comp}
 &\Omega^4\colon \
 f'
 =
 f
\end{align}
(where $e_j$ denotes unit vector along $x_j$). They can be written via multiplication by
\[ 
\mymathbb{1} + \epsilon N_j,
\] 
where
\begin{align*} 
&\Omega^0\colon \ N_j =
 0, \\
 &\Omega^1\colon \ N_j =
 \left( \begin{matrix}
 0 & \lambda_2 e_j^{\rm T} \\
 \lambda_1 e_j & 0
 \end{matrix}
 \right), \\
&\Omega^2\colon \ N_j =
 \left( \begin{matrix}
 0 & - \lambda_2 l_j \\
 \lambda_1 l_j & 0
 \end{matrix}
 \right), \\
 &\Omega^3\colon \ N_j =
 \left( \begin{matrix}
 0 & \lambda_1 e_j^{\rm T} \\
 \lambda_2 e_j & 0
 \end{matrix}
 \right), \\
 &\Omega^4\colon \ N_j = 0.
\end{align*}

\subsection[List of all generators rho\_p]{List of all generators $\boldsymbol{\rho_p}$}\label{generrotboost}

If we denote, in general,
\[ 
 \rho_p = S_j \ \text{and} \ N_j \ \text{on components of} \ \text{$p$-forms} ,
\] 
then the full list of generators on (suitably parametrized components of) $p$-forms explicitly reads
\begin{align} 
 \label{lorgalcar0comp}
 &\rho_0\colon \ S_j = 0, \qquad N_j = 0; \\
 \label{lorgalcar1comp}
 &\rho_1\colon \
 S_j =
 \left( \begin{matrix}
 0 & 0 \\
 0 & l_j
 \end{matrix}
 \right),\qquad
 N_j =
 \left( \begin{matrix}
 0 & \lambda_2 e_j^{\rm T} \\
 \lambda_1 e_j & 0
 \end{matrix}
 \right); \\
 \label{lorgalcar2comp}
 &\rho_2\colon \
 S_j =
 \left( \begin{matrix}
 l_j & 0 \\
 0 & l_j
 \end{matrix}
 \right),\qquad
 N_j =
 \left( \begin{matrix}
 0 & - \lambda_2 l_j \\
 \lambda_1 l_j & 0
 \end{matrix}
 \right); \\
 \label{lorgalcar3comp}
&\rho_3\colon \
 S_j =
 \left( \begin{matrix}
 0 & 0 \\
 0 & l_j
 \end{matrix}
 \right),\qquad
 N_j =
 \left( \begin{matrix}
 0 & \lambda_1 e_j^{\rm T} \\
 \lambda_2 e_j & 0
 \end{matrix}
 \right); \\
 \label{lorgalcar4comp}
 &\rho_4\colon \ S_j = 0, \qquad N_j = 0.
\end{align}
Observe a simple rule, which can save some computations in the future:
\begin{enumerate}\itemsep=0pt
\item[--] Interchanging $(\lambda_1,\lambda_2) \leftrightarrow (\lambda_2,\lambda_1)$ leads to interchanging $\rho_1 \leftrightarrow \rho_3$
\end{enumerate}
or, as a formula,
\begin{gather} \label{rho1rho3andswaplamdba1lambda2}
 \rho_3 (\lambda_1,\lambda_2) = \rho_1 (\lambda_2,\lambda_1).
\end{gather}

\section[Intertwining operators a\_\{qp\}]{Intertwining operators $\boldsymbol{a_{qp}}$}
\label{intertwiningaqp}

\subsection{What intertwining operators we are interested in}
\label{intertwinminkgalcar}

On Minkowski, Galilei or Carroll spacetime, respectively, let
\[ 
 a_{qp} \colon \ \Omega^p \to \Omega^q
\] 
be a linear operator realized as a \emph{matrix} on \emph{components} of the forms.
(Notice the order of indices convention -- mapping from $p$ to $q$ is denoted as $a_{qp}$ rather than vice versa.)

Recall (see (\ref{columnsformsgalcarlor})) that the components of $p$-forms are parametrized,
for $p=0,\dots ,4$, either by single objects or by pairs.
Therefore, the matrices $a_{qp}$ are necessarily (constant) \emph{block} matrices of (block) dimensions
$1\times 1$, $1\times 2$, $2\times 1$ or $2\times 2$.
As an example, matrices $a_{12}$ and $a_{02}$ are of the form
\begin{gather} \label{a12a02}
 a_{12} =
 \left( \begin{matrix}
 \mathbf c^{\rm T} & \mathbf d^{\rm T} \\
 C & D
 \end{matrix}
 \right),
\qquad
 a_{02} =
 \left( \begin{matrix}
 \mathbf c^{\rm T} & \mathbf d^{\rm T}
 \end{matrix}
 \right)
\end{gather}
for $\mathbf c$, $\mathbf d$ (column) vectors (i.e., $\mathbf c^{\rm T}$, $\mathbf d^{\rm T}$ row vectors) and $C$, $D$ $3\times 3$ matrices.

Within the spaces $\Omega^p$ and $\Omega^q$,
the corresponding (Lorentz, Galilei or Carroll) transformations induce (as generators)
matrices $\rho_p$ and $\rho_q$, respectively,
acting on components of forms and explicitly given, for both rotations and boosts,
by (\ref{lorgalcar0comp})--(\ref{lorgalcar4comp}).

Now we want $a_{qp}$ to be \emph{intertwining operator} for the two representations.
So we require that standard commutative diagram holds:
\begin{equation}
\begin{CD} \label{intertwinptoq}
 \Omega^p @>a_{qp}>> \Omega^q \\
 @V{\rho_p}VV @VV{\rho_q}V \\
 \Omega^p @>>a_{qp}> \Omega^q \\
\end{CD}
\qquad \text{i.e.,} \qquad
 \rho_q \circ a_{qp} = a_{qp} \circ \rho_p.
\end{equation}
Thus, we are to solve equation (\ref{intertwinptoq}) for each $q,p=0,\dots , 4$
(and any $j$ hidden in $\rho_p$ and $\rho_q$).

Realize that, as is always the case for intertwining operators, solutions constitute
a \emph{vector space}, so that we are actually to find \emph{a basis} of the space of solutions.

As an example, for $a_{12}$ and $a_{02}$, we are to solve
\begin{gather*}
 \rho_1 \circ a_{12} = a_{12} \circ \rho_2,
\qquad
 \rho_0 \circ a_{02} = a_{02} \circ \rho_2
\end{gather*}
for
\begin{enumerate}\itemsep=0pt
\item[--] matrices $a_{12}$ and $a_{02}$ (i.e., for unknown $\mathbf c$, $\mathbf d$, $C$ and $D$ in (\ref{a12a02})),
\item[--]$\rho_{0}$, $\rho_{1}$ and $\rho_{2}$ displayed in (\ref{lorgalcar0comp}), (\ref{lorgalcar1comp}) and (\ref{lorgalcar2comp}).
\end{enumerate}
Performing this for \emph{rotation} generators $\rho_{0}$, $\rho_{1}$ and $\rho_{2}$ \emph{alone}
(i.e., only for $S_j$ operators from (\ref{lorgalcar0comp}), (\ref{lorgalcar1comp}) and (\ref{lorgalcar2comp})), we get
\begin{gather} \label{a12a02porot}
 a_{12} =
 \left( \begin{matrix}
 \mathbf 0^{\rm T} & \mathbf 0^{\rm T} \\
 k_1 \mymathbb{1} & k_2 \mymathbb{1}
 \end{matrix}
 \right),
\qquad
 a_{02} = 0 \equiv
 \left( \begin{matrix}
 \mathbf 0^{\rm T} & \mathbf 0^{\rm T}
 \end{matrix}
 \right).
\end{gather}
So, $a_{02}$ is completely fixed by rotations alone
(it \emph{vanishes} -- there is \emph{no non-vanishing} intertwining~$a_{02}$
 \emph{irrespective} of the spacetime under consideration) and in $a_{12}$
still two free constants survive.

Repeating the same procedure with $a_{12}$ (already with ansatz (\ref{a12a02porot}))
for \emph{boost} generators $\rho_{1}$ and $\rho_{2}$
 we get the following restrictions on $k_1$, $k_2$ (see the details in Appendix \ref{detailssolving}):
\[ 
 k_1\lambda_2 = k_2\lambda_2 = k_2\lambda_1 = 0
\] 
(notice that they already depend on $\lambda_1$, $\lambda_2$,
 i.e., on whether we speak of Lorentz, Galilei or Carroll case).
This should be analyzed for all three cases (\ref{lorlambdas})--(\ref{carlambdas})
separately and the result is that the \emph{only non-vanishing} operator is
(constant multiple of)
\[ 
 a_{12} =
 \left( \begin{matrix}
 \mathbf 0^{\rm T} & \mathbf 0^{\rm T} \\
\mymathbb{1} & \mymathbb{0}
 \end{matrix}
 \right)
\qquad
 \text{for \emph{Carroll} spacetime}.
\] 
When translated back into the language of complete forms (rather than just their component column),
it produces the following (intertwining) mapping $\Omega^2 \to \Omega^1$:
\[ 
 a_{12}\colon
\
 {\rm d}t \wedge \mathbf a \cdot {\rm d}\mathbf r + \mathbf b \cdot {\rm d}\mathbf S
 \mapsto
 \mathbf a \cdot {\rm d}\mathbf r
\qquad
 \text{for \emph{Carroll} spacetime}.
\] 
In what follows we perform similar calculation for \emph{all} $a_{qp}$
(and \emph{all three} spacetimes).

\subsection[General structure of all a\_\{qp\} (no invariance, yet)]{General structure of all $\boldsymbol{a_{qp}}$ (no invariance, yet)}
\label{intertwinminkgalcargeneral}

A priori, just from the parametrization structure (\ref{columnsformsgalcarlor})
of spaces $\Omega^q$ and $\Omega^p$
(with no intertwining properties, yet) the operators $a_{qp}$
\begin{alignat*}{6} 
 &a_{00}\qquad& &
 a_{01}\qquad& &
 a_{02}\qquad& &
 a_{03}\qquad& &
 a_{04}& \\
 &a_{10}\qquad& &
 a_{11}\qquad& &
 a_{12}\qquad& &
 a_{13}\qquad& &
 a_{14}& \\
 &a_{20}\qquad& &
 a_{21}\qquad& &
 a_{22}\qquad& &
 a_{23\qquad}& &
 a_{24}& \\
 &a_{30}\qquad& &
 a_{31}\qquad& &
 a_{32}\qquad& &
 a_{33}\qquad& &
 a_{34}& \\
 &a_{40}\qquad& &
 a_{41}\qquad& &
 a_{42}\qquad& &
 a_{43}\qquad& &
 a_{44}&
\end{alignat*}
should be parametrized as follows
(we already encountered $a_{12}$ and $a_{02}$ in (\ref{a12a02})):
\begin{alignat}{6} 
 \label{a00toa04general2}
 &\quad k \qquad&
 &\left( \begin{matrix}
 k & \mathbf c^{\rm T}
 \end{matrix}
 \right) \qquad&
 & \left( \begin{matrix}
 \mathbf c^{\rm T} & \mathbf d^{\rm T}
 \end{matrix}
 \right) \qquad&
 &\left( \begin{matrix}
 k & \mathbf c^{\rm T}
 \end{matrix}
 \right) \qquad&
 & \quad k &\\
& \label{a10toa14general2}
 \left( \begin{matrix}
 k \\
 \mathbf c
 \end{matrix}
 \right) \qquad&
 &\left( \begin{matrix}
 k & \mathbf c^{\rm T} \\
 \mathbf d & C
 \end{matrix}
 \right) \qquad&
 &\left( \begin{matrix}
 \mathbf c^{\rm T} & \mathbf d^{\rm T} \\
 C & D
 \end{matrix}
 \right) \qquad&
 &\left( \begin{matrix}
 k & \mathbf c^{\rm T} \\
 \mathbf d & C
 \end{matrix}
 \right) \qquad&
 &\left( \begin{matrix}
 k \\
 \mathbf c
 \end{matrix}
 \right) &\\
 \label{a20toa24general2}
 &\left( \begin{matrix}
 \mathbf c \\
 \mathbf d
 \end{matrix}
 \right) \qquad&
 &\left( \begin{matrix}
 \mathbf c & C \\
 \mathbf d & D
 \end{matrix}
 \right) \qquad&
 &\left( \begin{matrix}
 C & D \\
 E & F
 \end{matrix}
 \right) \qquad&
 &\left( \begin{matrix}
 \mathbf c & C \\
 \mathbf d & D
 \end{matrix}
 \right) \qquad&
 &\left( \begin{matrix}
 \mathbf c \\
 \mathbf d
 \end{matrix}
 \right) &\\
 \label{a30toa34general2}
 &\left( \begin{matrix}
 k \\
 \mathbf c
 \end{matrix}
 \right) \qquad&
 &\left( \begin{matrix}
 k & \mathbf c^{\rm T} \\
 \mathbf d & C
 \end{matrix}
 \right) \qquad&
 &\left( \begin{matrix}
 \mathbf c^{\rm T} & \mathbf d^{\rm T} \\
 C & D
 \end{matrix}
 \right) \qquad&
 &\left( \begin{matrix}
 k & \mathbf c^{\rm T} \\
 \mathbf d & C
 \end{matrix}
 \right) \qquad&
 &\left( \begin{matrix}
 k \\
 \mathbf c
 \end{matrix}
 \right)& \\
 \label{a40toa44general2}
 & \quad k \qquad&
 &\left( \begin{matrix}
 k & \mathbf c^{\rm T}
 \end{matrix}
 \right) \qquad&
 &\left( \begin{matrix}
 \mathbf c^{\rm T} & \mathbf d^{\rm T}
 \end{matrix}
 \right) \qquad&
 &\left( \begin{matrix}
 k & \mathbf c^{\rm T}
 \end{matrix}
 \right) \qquad&
 & \quad k.&
\end{alignat}
Notice the symmetry of the structure of the matrices:
The pattern (type of matrix) happens to be symmetric with respect to both
\begin{enumerate}\itemsep=0pt
\item[--] reflection across the \emph{central horizontal} axis and
\item[--] reflection across the \emph{central vertical} axis.
\end{enumerate}
This is just a simple consequence of the symmetry of parametrizations (\ref{columnsformsgalcarlor})
w.r.t.\ left-right reflection over the center. So there are altogether 9 \emph{types} of matrices $a_{qp}$.

\subsection[How a\_\{qp\} are restricted by rotation invariance alone]{How $\boldsymbol{a_{qp}}$ are restricted by rotation invariance alone}
\label{intertwinminkgalcargenplusrot}

After applying rotational invariance alone
(i.e., property (\ref{intertwinptoq}) with just rotation generators~$S_j$),
the matrices (\ref{a00toa04general2})--(\ref{a40toa44general2}) simplify a lot.
In short the simplification may be described as application of the following rules:
All
\begin{enumerate}\itemsep=0pt
\item[--] scalars remain intact (all scalars are rotation invariant),
\item[--] vectors become zero (no rotation invariant vectors except for zero),
\item[--] matrices become multiples of unity (no other rotation invariant matrices).
\end{enumerate}
What we get is the following, much simpler, parametrization:
\begin{alignat}{6}
 \label{a00toa04afterrot}
 &\quad k \qquad &
 &
 \left( \begin{matrix}
 k & \mathbf 0^{\rm T}
 \end{matrix}
 \right) \qquad&
 &
 \left( \begin{matrix}
 \mathbf 0^{\rm T} & \mathbf 0^{\rm T}
 \end{matrix}
 \right) \qquad &
 &
 \left( \begin{matrix}
 k & \mathbf 0^{\rm T}
 \end{matrix}
 \right) \qquad &
 &\quad k & \\
 \label{a10toa14afterrot}
 &\left( \begin{matrix}
 k \\
 \mathbf 0
 \end{matrix}
 \right) \qquad &
 &\left( \begin{matrix}
 k_1 & \mathbf 0^{\rm T} \\
 \mathbf 0 & k_2 \mymathbb{1}
 \end{matrix}
 \right) \qquad &
 &\left( \begin{matrix}
 \mathbf 0^{\rm T} & \mathbf 0^{\rm T} \\
 k_1 \mymathbb{1} & k_2 \mymathbb{1}
 \end{matrix}
 \right) \qquad &
 &\left( \begin{matrix}
 k_1 & \mathbf 0^{\rm T} \\
 \mathbf 0 & k_2 \mymathbb{1}
 \end{matrix}
 \right) \qquad &
 &\left( \begin{matrix}
 k \\
 \mathbf 0
 \end{matrix}
 \right) & \\
 \label{a20toa24afterrot}
 &\left( \begin{matrix}
 \mathbf 0 \\
 \mathbf 0
 \end{matrix}
 \right) \qquad &
 &\left( \begin{matrix}
 \mathbf 0 & k_1 \mymathbb{1} \\
 \mathbf 0 & k_2 \mymathbb{1}
 \end{matrix}
 \right) \qquad &
 &\left( \begin{matrix}
 k_1 \mymathbb{1} & k_2 \mymathbb{1} \\
 k_3 \mymathbb{1} & k_4 \mymathbb{1}
 \end{matrix}
 \right) \qquad &
 &\left( \begin{matrix}
 \mathbf 0 & k_1 \mymathbb{1} \\
 \mathbf 0 & k_2 \mymathbb{1}
 \end{matrix}
 \right) \qquad &
 &\left( \begin{matrix}
 \mathbf 0 \\
 \mathbf 0
 \end{matrix}
 \right) & \\
 \label{a30toa34afterrot}
 &\left( \begin{matrix}
 k \\
 \mathbf 0
 \end{matrix}
 \right) \qquad &
 &\left( \begin{matrix}
 k_1 & \mathbf 0^{\rm T} \\
 \mathbf 0 & k_2 \mymathbb{1}
 \end{matrix}
 \right) \qquad &
 &\left( \begin{matrix}
 \mathbf 0^{\rm T} & \mathbf 0^{\rm T} \\
 k_1 \mymathbb{1} & k_2 \mymathbb{1}
 \end{matrix}
 \right) \qquad &
 &\left( \begin{matrix}
 k_1 & \mathbf 0^{\rm T} \\
 \mathbf 0 & k_2 \mymathbb{1}
 \end{matrix}
 \right) \qquad &
 &\left( \begin{matrix}
 k \\
 \mathbf 0
 \end{matrix}
 \right) & \\
 \label{a40toa44afterrot}
 &\quad k \qquad &
 &
 \left( \begin{matrix}
 k & \mathbf 0^{\rm T}
 \end{matrix}
 \right) \qquad &
 &
 \left( \begin{matrix}
 \mathbf 0^{\rm T} & \mathbf 0^{\rm T}
 \end{matrix}
 \right) \qquad &
 &
 \left( \begin{matrix}
 k & \mathbf 0^{\rm T}
 \end{matrix}
 \right)\qquad &
 &\quad k.&
\end{alignat}

\subsection[Restrictions on a\_\{qp\} added by boosts]{Restrictions on $\boldsymbol{a_{qp}}$ added by boosts}
\label{intertwinminkgalcargenplusrotplusboosts}

There are further restrictions on operators $a_{qp}$ displayed in (\ref{a00toa04afterrot})--(\ref{a40toa44afterrot})
added by boosts. Since boost generators already depend on $(\lambda_1, \lambda_2)$,
these restrictions depend on the choice of particular spacetime.
So we get, in general, different set of operators $a_{qp}$ for different spacetimes.

An example of detailed computations is presented in Appendix \ref{detailssolving}.

Here we give the results (and comments).

\section[Results - non-trivial operators a\_\{qp\}]{Results -- non-trivial operators $\boldsymbol{a_{qp}}$}
\label{aqpminkgalcarnontrivial}

\subsection[Which operators a\_\{qp\} are trivial]{Which operators $\boldsymbol{a_{qp}}$ are trivial}
\label{intertwinminkgalcartrivial}

It is evident from the structure of (\ref{intertwinptoq}) that
\begin{enumerate}\itemsep=0pt
\item[--] \emph{zero} operator (for any pair of spaces, $\Omega^q \to \Omega^p$) as well as
\item[--] \emph{unit} operator within any space ($\hat 1\colon \Omega^p \to \Omega^p$)
\end{enumerate}
are solutions, i.e., intertwining operators.

Unit operators form part of those found by our computation;
so, for example, $a_{11}$ and $a_{33}$ from~(\ref{a11a33lorgalcar})
are in a sense trivial and they may be omitted from our further discussion.
(And they clearly work in \emph{any} spacetime.)

In what follows we summarize all \emph{non-trivial} intertwining operators
found in our systematic computations.
They depend, as we have seen, on particular spacetime.

\subsection[Non-trivial operators a\_\{qp\} on Minkowski spacetime]{Non-trivial operators $\boldsymbol{a_{qp}}$ on Minkowski spacetime}
\label{intertwinminknontrivial}

On Minkowski spacetime we have found the following non-trivial ope\-rators:
\begin{gather*} 
 a_{40} = a_{04} = 1, \qquad
 a_{31} = a_{13} = \left( \begin{matrix}
 1 & \mathbf 0^{\rm T} \\
 \mathbf 0 & \mymathbb{1}
 \end{matrix}
 \right), \qquad
 a_{22} = \left( \begin{matrix}
 \hphantom{-}\mymathbb{0} & \mymathbb{1} \\
 - \mymathbb{1} & \mymathbb{0}
 \end{matrix}
 \right)
\end{gather*}
(see (\ref{a13lor}) and (\ref{a31lor});
 we use brief notation $a_{22}$ ignoring its (trivial) \emph{unit} part).

From the matrices displayed above we can read-off how the five operators actually act
on corresponding differential forms:
\begin{align} 
 \label{a40onOmega0lor}
 &a_{40}\colon\ f \mapsto f{\rm d}t \wedge {\rm d}V, \\
 \label{a31onOmega0lor}
 &a_{31}\colon\ f {\rm d}t + \mathbf a \cdot {\rm d}\mathbf r
 \mapsto {\rm d}t \wedge \mathbf a \cdot {\rm d}\mathbf S + f{\rm d}V, \\
 \label{a22onOmega0lor}
 &a_{22}\colon\ {\rm d}t \wedge \mathbf a \cdot {\rm d}\mathbf r + \mathbf b \cdot {\rm d}\mathbf S
 \mapsto {\rm d}t \wedge \mathbf b \cdot {\rm d}\mathbf r - \mathbf a \cdot {\rm d}\mathbf S, \\
 \label{a13onOmega0lor}
 &a_{13}\colon\ {\rm d}t \wedge \mathbf a \cdot {\rm d}\mathbf S + f{\rm d}V
 \mapsto f{\rm d}t + \mathbf a \cdot {\rm d}\mathbf r, \\
 \label{a04onOmega0lor}
 &a_{04}\colon\ f{\rm d}t \wedge {\rm d}V
 \mapsto f.
\end{align}
Now if we compare this with the action of standard (Minkowski) \emph{Hodge star} operator
$* \equiv *_{\text{Min}}$ (see, e.g., \cite[Section~16.1]{Feckoangl}) on forms
\begin{gather} 
 \label{*MinkOmega0a}
 *f
 = f{\rm d}t \wedge {\rm d}V, \\
 \label{*MinkOmega1a}
 *(f{\rm d}t + \mathbf a \cdot {\rm d}\mathbf r) = {\rm d}t \wedge \mathbf a \cdot {\rm d}\mathbf S + f{\rm d}V, \\
 \label{*MinkOmega2a}
 *({\rm d}t \wedge \mathbf a \cdot {\rm d}\mathbf r + \mathbf b \cdot {\rm d}\mathbf S) = {\rm d}t \wedge \mathbf b \cdot {\rm d}\mathbf r - \mathbf a \cdot {\rm d}\mathbf S, \\
 \label{*MinkOmega3a}
 *({\rm d}t \wedge \mathbf a \cdot {\rm d}\mathbf S + f{\rm d}V) = f{\rm d}t + \mathbf a \cdot {\rm d}\mathbf r, \\
 \label{*MinkOmega4a}
 *(f{\rm d}t \wedge {\rm d}V) = -f,
\end{gather}
wee see that
\begin{enumerate}\itemsep=0pt
\item[--] we merely ``rediscovered'' \emph{standard} Hodge star in Minkowski spacetime,
\item[--] \emph{nothing more} than (a multiple of) the Hodge star is intertwining on forms, there.
\end{enumerate}
Invariance of the Hodge star is well known.
So what we \emph{actually} learned about the Minkowski case is that the Hodge star
is in fact \emph{the only} relevant operator with the property, there.
(And this is fairly plausible result; otherwise we certainly met the other operators
 in important physics equations.)

As we will see in the next two paragraphs, the situation in Galilei as well as Carroll
spacetimes turns out to be \emph{refreshingly different}.

\subsection[Non-trivial operators a\_\{qp\} on Galilei spacetime]{Non-trivial operators $\boldsymbol{a_{qp}}$ on Galilei spacetime}
\label{intertwingalnontrivial}

On \emph{Galilei} spacetime we have found quite a lot of non-trivial operators.

First, the following five \emph{analogs} of Minkowski ones:
\begin{gather} 
 \label{a40gal2}
 a_{40} = a_{04} = 1, \\
 \label{a31gal2}
 a_{31} = \left( \begin{matrix}
 0 & \mathbf 0^{\rm T} \\
 \mathbf 0 & \mymathbb{1}
 \end{matrix}
 \right), \\
 \label{a22gal2}
 a_{22} = \left( \begin{matrix}
 \mymathbb{0} & \mymathbb{1} \\
 \mymathbb{0} & \mymathbb{0}
 \end{matrix}
 \right), \\
 \label{a13gal2}
 a_{13} = \left( \begin{matrix}
 1 & \mathbf 0^{\rm T} \\
 \mathbf 0 & \mymathbb{0}
 \end{matrix}
 \right)
\end{gather}
(see (\ref{a13gal}) and (\ref{a31gal})).
On corresponding differential forms:
\begin{align} 
 \label{a40onOmega0gal}
 &a_{40}\colon\ f \mapsto f{\rm d}t \wedge {\rm d}V, \\
 \label{a31onOmega1gal}
 &a_{31}\colon\ f{\rm d}t + \mathbf a \cdot {\rm d}\mathbf r
 \mapsto {\rm d}t \wedge \mathbf a \cdot {\rm d}\mathbf S, \\
 \label{a22onOmega2gal}
 &a_{22}\colon\ {\rm d}t \wedge \mathbf a \cdot {\rm d}\mathbf r + \mathbf b \cdot {\rm d}\mathbf S
 \mapsto {\rm d}t \wedge \mathbf b \cdot {\rm d}\mathbf r, \\
 \label{a13onOmega3gal}
 &a_{13}\colon\ {\rm d}t \wedge \mathbf a \cdot {\rm d}\mathbf S + f{\rm d}V
 \mapsto f{\rm d}t, \\
 \label{a04onOmega4gal}
 &a_{04}\colon\ f{\rm d}t \wedge {\rm d}V
 \mapsto f.
\end{align}
In \emph{Minkowski} case, see (\ref{a40onOmega0lor})--(\ref{a04onOmega0lor}), we realized that
\emph{this type} of operators was nothing but the \emph{Hodge star} acting on various degrees of forms,
see (\ref{*MinkOmega0a})--(\ref{*MinkOmega4a}).
Here this is, in a sense, a similar story. One can introduce ``Galilei Hodge star'' on Galilei spacetime
(see Appendix \ref{app:galcarminhodgefromFecko} and for more details, including motivation, see \cite{FeckoGalCarHodge})
and it turns out that (\ref{a40onOmega0gal})--(\ref{a04onOmega4gal})
just reproduce the formulas valid for the Galilei Hodge star.
(Actually there are as many as two possibilities for the Galilei Hodge described in \cite{FeckoGalCarHodge}
 and the last operator, $a_{04}$, reproduces one of them (the other gives zero.)
So we can say again, like we did at the end of Section \ref{intertwinminknontrivial},
that in (\ref{a40onOmega0gal})--(\ref{a04onOmega4gal}) we ``merely rediscovered''
\emph{Galilei} Hodge star on Galilei spacetime.

Now there are, however, in addition to (\ref{a40gal2})--(\ref{a13gal2}), operators with \emph{no analogy}
to Minkowski situation. They act as follows:
\[ 
 \Omega^0 \overset {a_{_{10}}} \to
 \Omega^1 \overset {a_{21}} \to
 \Omega^2 \overset {a_{32}} \to
 \Omega^3 \overset {a_{43}} \to
 \Omega^4.
\] 
Here
\begin{gather} 
 a_{10} = \left( \begin{matrix}
 1 \\
 \mathbf 0
 \end{matrix}
 \right), \qquad
 a_{21} = \left( \begin{matrix}
 \mathbf 0 & \mymathbb{1} \\
 \mathbf 0 & \mymathbb{0}
 \end{matrix}
 \right), \qquad
 \label{a32gal2}
 a_{32} = \left( \begin{matrix}
 \mathbf 0^{\rm T} & \mathbf 0^{\rm T} \\
 \mymathbb{0} & \mymathbb{1}
 \end{matrix}
 \right), \qquad
 a_{43} = \left( \begin{matrix}
 1 & \mathbf 0^{\rm T}
 \end{matrix}
 \right).
\end{gather}
So their actual action on corresponding differential forms reads
\begin{align*} 
& a_{10}\colon \ f \mapsto f{\rm d}t, \\
 &a_{21}\colon \ f{\rm d}t + \mathbf a \cdot {\rm d}\mathbf r
 \mapsto {\rm d}t \wedge \mathbf a \cdot {\rm d}\mathbf r, \\
& a_{32}\colon \ {\rm d}t \wedge \mathbf a \cdot {\rm d}\mathbf r + \mathbf b \cdot {\rm d}\mathbf S
 \mapsto {\rm d}t \wedge \mathbf b \cdot {\rm d}\mathbf S, \\
 &a_{43}\colon \ {\rm d}t \wedge \mathbf a \cdot {\rm d}\mathbf S + f{\rm d}V
 \mapsto f{\rm d}t \wedge {\rm d}V.
\end{align*}
We see that the rule how these operators work is nothing but application of \emph{exterior product}
\begin{gather} \label{apg=dtwedge}
 (\dots ) \mapsto {\rm d}t \wedge (\dots )
\end{gather}
on the form under consideration. This is clearly
\begin{enumerate}\itemsep=0pt
\item[--] an operator of degree $+1$ and it is also,
\item[--] \emph{Galilei-invariant} operation on forms,
\end{enumerate}
since
 \[
 {\rm d}t' = {\rm d}t
\]
for Galilei boosts (\ref{boostgalt}) and (\ref{boostgalr}). (Speak nothing of rotations and translations.)
Finally, it should be noted that there were still two more operators found, namely
\begin{gather*} 
 a_{03} = \left( \begin{matrix}
 1 & \mathbf 0^{\rm T}
 \end{matrix}
 \right), \qquad
 a_{14} = \left( \begin{matrix}
 1 \\
 \mathbf 0
 \end{matrix}
 \right),
\end{gather*}
i.e., with action on forms
\begin{align*} 
 &a_{03}\colon \ {\rm d}t \wedge \mathbf a \cdot {\rm d}\mathbf S + f{\rm d}V
 \mapsto f, \qquad
 a_{14}\colon\ f{\rm d}t \wedge {\rm d}V
 \mapsto f{\rm d}t.
\end{align*}
Both of them are, however, just \emph{compositions} of operators already described previously,
namely
\begin{gather*} 
 a_{03} = a_{04} \circ a_{43}, \qquad
 a_{14} = a_{10} \circ a_{04}.
\end{gather*}

\subsection[Non-trivial operators a\_\{qp\} on Carroll spacetime]{Non-trivial operators $\boldsymbol{a_{qp}}$ on Carroll spacetime}
\label{intertwincarnontrivial}

On \emph{Carroll} spacetime we have found quite a lot of non-trivial ope\-rators, too.

First, the following five \emph{analogs} of Lorentzian ones:
\begin{gather*} 
 a_{40} = a_{04} = 1, \qquad
 a_{31} = \left( \begin{matrix}
 1 & \mathbf 0^{\rm T} \\
 \mathbf 0 & \mymathbb{0}
 \end{matrix}
 \right), \qquad
 a_{22} = \left( \begin{matrix}
 \mymathbb{0} & \mymathbb{0} \\
 \mymathbb{1} & \mymathbb{0}
 \end{matrix}
 \right), \qquad
 a_{13} = \left( \begin{matrix}
 0 & \mathbf 0^{\rm T} \\
 \mathbf 0 & \mymathbb{1}
 \end{matrix}
 \right).
\end{gather*}
(see (\ref{a13car}) and (\ref{a31car})).
On corresponding differential forms:
\begin{gather} 
 \label{a40onOmega0car}
 a_{40}\colon \ f \mapsto f{\rm d}t \wedge {\rm d}V, \\
 \label{a31onOmega1car}
 a_{31}\colon\ f{\rm d}t + \mathbf a \cdot {\rm d}\mathbf r
 \mapsto f{\rm d}V, \\
 \label{a22onOmega2car}
 a_{22}\colon\ {\rm d}t \wedge \mathbf a \cdot {\rm d}\mathbf r + \mathbf b \cdot {\rm d}\mathbf S
 \mapsto \mathbf a \cdot {\rm d}\mathbf S, \\
 \label{a13onOmega3car}
 a_{13}\colon\ {\rm d}t \wedge \mathbf a \cdot {\rm d}\mathbf S + f{\rm d}V
 \mapsto \mathbf a \cdot {\rm d}\mathbf r, \\
 \label{a04onOmega4car}
 a_{04}\colon\ f{\rm d}t \wedge {\rm d}V
 \mapsto f.
\end{gather}
Here the story is similar to the Galilei case. One can introduce ``Carroll Hodge star'' in Carroll spacetime
(again see Appendix \ref{app:galcarminhodgefromFecko} and \cite{FeckoGalCarHodge})
and it turns out that (\ref{a40onOmega0car})--(\ref{a04onOmega4car})
just reproduce the formulas valid for the Carroll Hodge star.

So we can say again, like we did at the end of Sections \ref{intertwinminknontrivial} and \ref{intertwingalnontrivial},
that in (\ref{a40onOmega0car})--(\ref{a04onOmega4car}) we ``merely rediscovered''
\emph{Carroll} Hodge star in Carroll spacetime.

In addition, again, there are operators with \emph{no analogy} to Minkowski situation
(albeit with a \emph{strong} analogy to the \emph{Galilei} one). They act as follows:
\[
 \Omega^0 \overset {a_{_{01}}} \leftarrow
 \Omega^1 \overset {a_{12}} \leftarrow
 \Omega^2 \overset {a_{23}} \leftarrow
 \Omega^3 \overset {a_{34}} \leftarrow
 \Omega^4.
\]
Here
\begin{gather*} 
 a_{01} = \left( \begin{matrix}
 1 & \mathbf 0^{\rm T}
 \end{matrix}
 \right), \qquad
 a_{12} = \left( \begin{matrix}
 \mathbf 0^{\rm T} & \mathbf 0^{\rm T} \\
 \mymathbb{1} & \mymathbb{0}
 \end{matrix}
 \right), \qquad
 a_{23} = \left( \begin{matrix}
 \mathbf 0 & \mymathbb{0} \\
 \mathbf 0 & \mymathbb{1}
 \end{matrix}
 \right), \qquad
 a_{34} = \left( \begin{matrix}
 1 \\
 \mathbf 0
 \end{matrix}
 \right).
\end{gather*}
On corresponding differential forms it is
\begin{gather*} 
 a_{01}\colon \ f{\rm d}t + \mathbf a \cdot {\rm d}\mathbf r \mapsto f, \\
 a_{12}\colon\ {\rm d}t \wedge \mathbf a \cdot {\rm d}\mathbf r + \mathbf b \cdot {\rm d}\mathbf S
 \mapsto \mathbf a \cdot {\rm d}\mathbf r, \\
 a_{23}\colon\ {\rm d}t \wedge \mathbf a \cdot {\rm d}\mathbf S + f{\rm d}V
 \mapsto \mathbf a \cdot {\rm d}\mathbf S, \\
 a_{34}\colon\ f{\rm d}t \wedge {\rm d}V
 \mapsto f {\rm d}V.
\end{gather*}
We can see that the rule how these four operators work is nothing but application of \emph{interior product}
\begin{gather} \label{iVforV=partialt}
 (\dots) \mapsto i_{\partial_t} (\dots)
\end{gather}
on the form under consideration. It is clearly
\begin{enumerate}\itemsep=0pt
\item[--] an operator of degree $-1$ and,
\item[--] it is also \emph{Carroll-invariant} operation on forms,
\end{enumerate}
since
 \[ 
 \partial_{t'} = \partial_t
\] 
for Carroll boost (\ref{boostcarinft}) and (\ref{boostcarinfr}). (Speak nothing of rotations and translations.)

Finally, it should be noted that there were still two more operators found, namely
\begin{gather*} 
 a_{30} = \left( \begin{matrix}
 1 \\
 \mathbf 0
 \end{matrix}
 \right), \qquad
 a_{41} = \left( \begin{matrix}
 1 & \mathbf 0^{\rm T}
 \end{matrix}
 \right),
\end{gather*}
i.e., with action on forms
\begin{gather*} 
 a_{30}\colon\ f
 \mapsto f{\rm d}V, \qquad
 a_{41}\colon\ f{\rm d}t + \mathbf a \cdot {\rm d}\mathbf r
 \mapsto
 f{\rm d}t \wedge {\rm d}V.
\end{gather*}
Both of them are, however, just \emph{compositions} of operators already described previously,
namely
\begin{gather*} 
 a_{30} = a_{34} \circ a_{40}, \qquad
 a_{41} = a_{40} \circ a_{01}.
\end{gather*}

\subsection[A note on the degree pm 1 operators]{A note on the degree $\boldsymbol{\pm 1}$ operators}
\label{degpm1operators}

In Galilean and Carrollian spacetimes,
invariant tensor fields is a standard knowledge.

In Galilei case, say, two of them are, within our notation (in adapted coordinates), tensors
\[ 
 \delta^{ij}\partial_i \otimes \partial_j \in \mathcal T^2_0,
\qquad
 {\rm d}t \in \mathcal T^0_1.
\] 

In Carroll case we have similarly
\[ 
 \delta_{ij}{\rm d}x^i \otimes {\rm d}x^j \in \mathcal T^0_2,
\qquad
 \partial_t \in \mathcal T^1_0.
\] 
For example, for the Galilean case, we can already see them, to credit just some fathers-founders,
in (differently looking) component presentations as
\begin{alignat*}{3} 
 &g^{\alpha \beta}t_\beta = 0
\qquad&&
 \text{in 1963 paper \cite{Trautman1963},}& \\
 &h^{ab}t_b = 0
\qquad&&
 \text{in 1966 paper \cite{Trautman1966},}& \\
& \gamma^{\alpha \beta}\psi_{\beta} = 0
\qquad&&
 \text{in 1972 paper \cite{Kunzle1972},}&
\end{alignat*}
(they all describe the same equation).

On the other side, among the recent occurrence,
we can mention notation and terminology
\begin{alignat}{3} 
 &\text{Galilean}\colon\ && \gamma\quad\text{spatial cometric},& \nonumber\\
 \label{galclock2}
 &&& \tau\quad\text{clock one-form};& \\
 &\text{Carrollian}\colon\ && h\quad\text{spatial metric},& \nonumber\\
 \label{carvector2}
&&&\xi\quad\text{Carrollian vector field},&
\end{alignat}
with{\samepage
\[
 \gamma ( \tau, \ \cdot \ ) = 0,
\qquad
 h ( \xi, \ \cdot \ ) = 0
\]
from 2020 in \cite{Figueroa2020}.}

Now a simple observation is that, just being aware of tensor fields
$\tau$ and $\xi$ from (\ref{galclock2}) and~(\ref{carvector2}),
it should be clear from the very beginning
that two \emph{invariant} degree $\pm 1$ \emph{operators} are available,
\[
 \tau \wedge (\dots) \quad \text{on} \ \Omega \ (\text{Galilei}),
\qquad
 i_\xi (\dots) \quad \text{on} \ \Omega \ (\text{Carroll}).
\]
Notice that both may be regarded as \emph{differentials} in corresponding de Rham complex.

Also notice that they have \emph{no counterpart} for forms on \emph{Lorentzian} spacetimes
since there is neither invariant one-form nor invariant vector field, there.
We identify these interpretations of our ``degree $\pm 1$ operators''
in (\ref{apg=dtwedge}) and (\ref{iVforV=partialt}).

\subsection{Invariant operators and invariant subspaces}
\label{whymoregalileancarrollian}

In Galilei and Carroll spacetimes, some features of resulting invariant operators
differ from those in Minkowski spacetime. Namely
\begin{enumerate}\itemsep=0pt
\item[--] analogs of Hodge star happen to be degenerate,
\item[--] there are operators connecting spaces which were not connected before.
\end{enumerate}
There is a simple \emph{formal} reason for these features could appear.
Namely, the re\-pre\-sentations of Galilei and Carroll groups on components of forms
possess \emph{nontrivial invariant} subspaces.

Due to Schur's lemma,
any intertwining operator between two \emph{irreducible} representations
is either zero or isomorphism.
So, for irreducible representations, \emph{non-zero} intertwining operators are necessarily isomorphisms,
and therefore they operate between spaces of equal dimensions.

Now representations of Lorentz group on the space of components of $p$-forms \emph{are irreducible}.
Indeed, we see directly from explicit formulas
(namely from (\ref{columnsformsgalcarlor}) and (\ref{boost0comp})--(\ref{boost4comp}))
that the space is

\begin{enumerate}\itemsep=0pt
\item[--] either 1-dimensional (0-forms and 4-forms),
\item[--] or spanned by \emph{pairs} ($(f,\mathbf a)$ or $(\mathbf a, \mathbf b)$),
\item[--] the two elements within the pair are \emph{independent} for rotations,
\item[--] the two elements within the pair \emph{need one another} for boosts,
\item[--] so for rotations \emph{and boosts} they \emph{need one another},
\item[--] so the \emph{whole} representation space is needed,
\item[--] so the representation is irreducible.
\end{enumerate}
Dimensionality of spaces of components of $p$-forms for $p=0,\dots ,4$ is
\[
 1 \qquad4 \qquad 6 \qquad4 \qquad 1
\]
so, just because of these \emph{dimensions},
non-trivial intertwining operators are only possible between spaces
$\Omega^0 \leftrightarrow \Omega^4$, $\Omega^1 \leftrightarrow \Omega^3$ and within each space, $\Omega^p \to \Omega^p$.
This is exactly used by the (Minkowski) Hodge star (the last possibility for $p=2$).
And, as an example, non-zero operator $\Omega^2 \to \Omega^3$ \emph{cannot} be intertwining.

For Galilei and Carroll spacetimes, the situation is different.

There \emph{are}, contrary to Minkowski case,
\emph{nontrivial invariant} sub-spaces
not only for rotations but also for Galilei (as well as Carroll) \emph{boosts}.

Namely formulas for infinitesimal rotations and Galilei/Carroll boosts,
based on generators listed in Section \ref{generrotboost},
readily show that, in our language for components of $p$-forms (\ref{columnsformsgalcarlor}),
there are invariant subspaces spanned on elements
\begin{alignat}{3} \label{columnsgalinvsubspaces}
 &\text{Galilei case}\colon
\ &&
 f,
\qquad
 \left( \begin{matrix}
 f \\
 \mathbf 0
 \end{matrix}
 \right),
\qquad
 \left( \begin{matrix}
 \mathbf a \\
 \mathbf 0
 \end{matrix}
 \right),
\qquad
 \left( \begin{matrix}
 0 \\
 \mathbf a
 \end{matrix}
 \right),
\qquad
 f,&\\
 &\text{Carroll case}\colon \ &&
 f,
\qquad
 \left( \begin{matrix}
 0 \\
 \mathbf a
 \end{matrix}
 \right),
\qquad
 \left( \begin{matrix}
 \mathbf 0 \\
 \mathbf b
 \end{matrix}
 \right),
\qquad
 \left( \begin{matrix}
 f \\
 \mathbf 0
 \end{matrix}
 \right),
\qquad
 f.&\nonumber
\end{alignat} 
So, we have \emph{non-trivial} invariant subspaces in $\Omega^1$, $\Omega^2$ and $\Omega^3$ and, therefore,
representations $\rho_1$, $\rho_2$ and $\rho_3$ are not irreducible.

Then, nothing (so simple as in the Lorentz case) excludes, e.g., a \emph{non-zero intertwining} operator
$\Omega^2 \to \Omega^3$. And indeed, we have found, in Galilei case, the operator
\[ 
 a_{32} \colon \
 \left( \begin{matrix}
 \mathbf a \\
 \mathbf b
 \end{matrix}
 \right)
 \mapsto
 \left( \begin{matrix}
 0 \\
 \mathbf b
 \end{matrix}
 \right)
\] 
(see (\ref{a32gal2})). Notice that
\[ 
 \Ker a_{32} \leftrightarrow \left( \begin{matrix}
 \mathbf a \\
 \mathbf 0
 \end{matrix}
 \right)
\qquad \text{and} \qquad
 \im a_{32} \leftrightarrow \left( \begin{matrix}
 0 \\
 \mathbf a
 \end{matrix}
 \right).
\] 
Kernel and image spaces of \emph{any} intertwining operator are known to be invariant subspaces,
and here we see that they coincide with particular invariant subspaces displayed in (\ref{columnsgalinvsubspaces}).

Similarly, the new Galilei Hodge star (acting on 1-forms)
\[ 
 a_{31} \colon \
 \left( \begin{matrix}
 f \\
 \mathbf a
 \end{matrix}
 \right)
 \mapsto
 \left( \begin{matrix}
 0 \\
 \mathbf a
 \end{matrix}
 \right)
\] 
has no reason to be isomorphism. And it indeed seized its chance:
\[ 
 \Ker a_{31} \leftrightarrow \left( \begin{matrix}
 f \\
 \mathbf 0
 \end{matrix}
 \right)
\qquad \text{and}\qquad
 \im a_{31} \leftrightarrow \left( \begin{matrix}
 0 \\
 \mathbf a
 \end{matrix}
 \right).
\] 
Again, they coincide with particular invariant subspaces displayed in (\ref{columnsgalinvsubspaces}).
Similar facts are true for the new Carroll Hodge star.

\section{Summary and conclusions}
\label{summaryconclusions}

Hodge star operator plays a key role in mathematical physics,
in particular it is frequently used on Lorentzian spacetimes.

A lot of research is recently done on \emph{Galilean} and \emph{Carrollian} spacetimes.
It is therefore natural to address the question about the use of the operator there.
We see immediately that it cannot be constructed in a straightforward way,
since the Hodge star is (by definition) associated with metric tensor
and there is no (canonical) metric tensor on the two spacetimes.

One can, however, look for potentially useful \emph{analogs} of the star operator.

In \cite{FeckoGalCarHodge}, this is achieved by proper modification of the standard formula
(making use of suitable invariant tensors). Results are summarized in Appendix \ref{app:galcarminhodgefromFecko}.

Here we follow another idea.
The search is based on observation of \emph{intertwining} (equivariance) property of the (original) Hodge
(in the sense explained in the text).
So we address the question of finding intertwining operators from $p$-forms to $q$-forms
on $(1+3)$-dimensional Galilean/Carrollian spacetimes.

The problem is first solved in Sections \ref{FormsGalCarMink}--\ref{aqpminkgalcarnontrivial}
for \emph{Galilei} and \emph{Carroll} spacetimes.

Then it is explained in Appendices \ref{galileicarrollvector}--\ref{spacetimeHodge}
why these computations actually guarantee that the corresponding results
also apply on general \emph{Galilean} and \emph{Carrollian} spacetimes
(and what small changes are needed).

What we found may be summarized as follows:
\begin{enumerate}\itemsep=0pt
\item On \emph{Lorentzian} spacetime, the Hodge star is \emph{the only} intertwining operator.
\item On \emph{Galilean} spacetime, there is
\begin{itemize}\itemsep=0pt
\item[--] Galilean Hodge star from \cite{FeckoGalCarHodge} plus,
\item[--] degree $+1$ operator $\alpha \mapsto \xi \wedge \alpha$.
 \end{itemize}
 \item On \emph{Carrollian} spacetime, there is
\begin{itemize}\itemsep=0pt
\item[--] Carrollian Hodge star from \cite{FeckoGalCarHodge} plus,
\item[--] degree $-1$ operator $\alpha \mapsto i_{\tilde \xi} \wedge \alpha$.
 \end{itemize}
\end{enumerate}

So these (algebraic) operators, when combined with exterior derivative $d$,
which \emph{is known} to be intertwining (differential) operator as well,
lead to systems of \emph{locally} Lorentz/Galilei/Carroll \emph{invariant equations}.
As an example, equations
\[ 
 {\rm d}*F= - J, \qquad {\rm d}F=0
\] 
(mentioned in Section \ref{intro})
correspond to Lorentzian/Galilean/Carrollian electrodynamics
depending on \emph{which particular} Hodge star is used; see more in~\cite{FeckoGalCarHodge}).

\appendix

\section[Details of solving equations for (some) a\_\{qp\}]{Details of solving equations for (some) $\boldsymbol{a_{qp}}$}
\label{detailssolving}

Here we provide details of solving equations (\ref{intertwinptoq}) for $a_{qp}$.
Not all of them (to save space), just a small sample to see a typical computation.

As mentioned in Section \ref{intertwinminkgalcargenplusrot},
the first step is taking \emph{rotations} generators $S_j$ alone.
This leads to simplification (\ref{a00toa04afterrot})--(\ref{a40toa44afterrot}).
Then we repeat, starting already with these simplified expressions,
the same step with \emph{boosts} generators $N_j$ alone.
Since they already depend on $(\lambda_1,\lambda_2)$,
results become dependent on particular spacetime.

Due to the symmetry of the pattern mentioned in Section~\ref{intertwinminkgalcargeneral}
we can often compute several operators at once
(perform a single computation valid for a particular group of operators,
 namely for the ``orbit'' of the symmetry of the pattern).

\emph{Nonzero} results (including \emph{unit} operators)
are highlighted by being $\boxed{\text{boxed}}$.

\subsection[Operators a\_\{11\}, a\_\{33\}, a\_\{13\} and a\_\{31\}]{Operators $\boldsymbol{a_{11}}$, $\boldsymbol{a_{33}}$, $\boldsymbol{a_{13}}$ and $\boldsymbol{a_{31}}$}
\label{a11a3310a13a31lorgalcar}

For $a_{11}$, $a_{33}$, $a_{13}$ and $a_{31}$, the equations read
\begin{alignat*} {3}
 &\rho_1 \circ a_{11} = a_{11} \circ \rho_1,\qquad && \rho_1 \circ a_{13} = a_{13} \circ \rho_3,& \\
 &\rho_3 \circ a_{33} = a_{33} \circ \rho_3, \qquad&& \rho_3 \circ a_{31} = a_{31} \circ \rho_1.&
\end{alignat*}
From this we see that
\begin{enumerate}\itemsep=0pt
\item[--]we only need to really compute $a_{11}(\lambda_1,\lambda_2)$ and $a_{13}(\lambda_1,\lambda_2)$,
\item[--]$a_{33}$ may be produced from $a_{11}$ by swapping $\lambda_1 \leftrightarrow \lambda_2$ (see (\ref{rho1rho3andswaplamdba1lambda2})),
\item[--]$a_{31}$ may be produced from $a_{13}$ by swapping $\lambda_1 \leftrightarrow \lambda_2$ (see (\ref{rho1rho3andswaplamdba1lambda2})).
\end{enumerate}
The equation for $a_{11}$ explicitly reads
\[ 
 \left( \begin{matrix}
 0 & \lambda_2 e_j^{\rm T} \\
 \lambda_1 e_j & 0
 \end{matrix}
 \right)
 \left( \begin{matrix}
 k_1 & \mathbf 0^{\rm T} \\
 \mathbf 0 & k_2 \mymathbb{1}
 \end{matrix}
 \right)
 =
 \left( \begin{matrix}
 k_1 & \mathbf 0^{\rm T} \\
 \mathbf 0 & k_2 \mymathbb{1}
 \end{matrix}
 \right)
 \left( \begin{matrix}
 0 & \lambda_2 e_j^{\rm T} \\
 \lambda_1 e_j & 0
 \end{matrix}
 \right),
\] 
i.e.,
\begin{equation} \label{a11lorgalcarexplicit2}
 k_2\lambda_2 = k_1\lambda_2,
 \qquad
 k_2\lambda_1 = k_1\lambda_1.
\end{equation}
Solution is
\[ 
 k_1 = k_2
\qquad
 \text{for all three spacetimes}.
\] 
And since the system (\ref{a11lorgalcarexplicit2}) remains intact for swapping $\lambda_1 \leftrightarrow \lambda_2$,
we get $a_{11}=a_{33}$.

The equation for $a_{13}$ explicitly reads
\[ 
 \left( \begin{matrix}
 0 & \lambda_2 e_j^{\rm T} \\
 \lambda_1 e_j & 0
 \end{matrix}
 \right)
 \left( \begin{matrix}
 k_1 & \mathbf 0^{\rm T} \\
 \mathbf 0 & k_2 \mymathbb{1}
 \end{matrix}
 \right)
 =
 \left( \begin{matrix}
 k_1 & \mathbf 0^{\rm T} \\
 \mathbf 0 & k_2 \mymathbb{1}
 \end{matrix}
 \right)
 \left( \begin{matrix}
 0 & \lambda_1 e_j^{\rm T} \\
 \lambda_2 e_j & 0
 \end{matrix}
 \right),
\] 
i.e.,
\begin{equation} \label{a13lorgalcarexplicit2}
 k_2\lambda_2 = k_1\lambda_1.
\end{equation}
Here we have for $a_{13}$
\begin{enumerate}\itemsep=0pt
\item[--]$\lambda_1=\lambda_2=1$ (Lorentz) $\Rightarrow$ ($k_1=k_2=$ arbitrary),
\item[--] $\lambda_1=0$ (Galilei) $\Rightarrow$ $k_2=0$ ($k_1$ arbitrary),
\item[--] $\lambda_2=0$ (Carroll) $\Rightarrow$ $k_1=0$ ($k_2$ arbitrary).
\end{enumerate}
Swapping $\lambda_1 \leftrightarrow \lambda_2$ in (\ref{a13lorgalcarexplicit2}) produces swapping
$k_1 \leftrightarrow k_2$ for solutions.
So it produces as $a_{31}$ swapping Galilei $\leftrightarrow$ Carroll for the above conclusion for $a_{13}$.

Altogether we come to conclusion:
\begin{gather} \label{a11a33lorgalcar}
 \boxed{
 a_{11}=a_{33}} =
 \left( \begin{matrix}
 1 & \mathbf 0^{\rm T} \\
 \mathbf 0 & \mymathbb{1}
 \end{matrix}
 \right),
\qquad
 \boxed{\text{Lorentz, Galilei, Carroll}}
\\
 \label{a13lor}
\boxed{a_{13}}
 = \left( \begin{matrix}
 1 & \mathbf 0^{\rm T} \\
 \mathbf 0 & \mymathbb{1}
 \end{matrix}
 \right)
\qquad
 \boxed{\text{Lorentz}} \\
 \label{a13gal}
 \phantom{\boxed{a_{13}}}{} = \left( \begin{matrix}
 1 & \mathbf 0^{\rm T} \\
 \mathbf 0 & \mymathbb{0}
 \end{matrix}
 \right)
\qquad
 \boxed{\text{Galilei}} \\
 \label{a13car}
 \phantom{\boxed{a_{13}}}{} = \left( \begin{matrix}
 0 & \mathbf 0^{\rm T} \\
 \mathbf 0 & \mymathbb{1}
 \end{matrix}
 \right),
\qquad
 \boxed{\text{Carroll}}
\\
 \label{a31lor}
 \boxed{a_{31}}
 = \left( \begin{matrix}
 1 & \mathbf 0^{\rm T} \\
 \mathbf 0 & \mymathbb{1}
 \end{matrix}
 \right)
\qquad
 \boxed{\text{Lorentz}} \\
 \label{a31car}
 \phantom{\boxed{a_{13}}}{} = \left( \begin{matrix}
 1 & \mathbf 0^{\rm T} \\
 \mathbf 0 & \mymathbb{0}
 \end{matrix}
 \right)
\qquad
 \boxed{\text{Carroll}} \\
 \label{a31gal}
 \phantom{\boxed{a_{13}}}{} = \left( \begin{matrix}
 0 & \mathbf 0^{\rm T} \\
 \mathbf 0 & \mymathbb{1}
 \end{matrix}
 \right).
\qquad
 \boxed{\text{Galilei}}
\end{gather}

\section{Galilean and Carrollian Hodge star operators}\label{app:galcarminhodgefromFecko}

In \cite{FeckoGalCarHodge}, Galilean and Carrollian ``Hodge star'' operators
were proposed via appropriate modification of the standard Hodge operator formula.
In order to compare the results obtained in this way with the intertwining operators
discussed in this paper, we list the results from~\cite{FeckoGalCarHodge}, here.
Just for the sake of completeness we start with the ``standard'' (Minkowski) Hodge star
(see \mbox{\cite[Section~16.1]{Feckoangl}}).

\medskip

\noindent
\emph{Minkowski} Hodge star:
\begin{gather*} 
 *f = f{\rm d}t \wedge {\rm d}V, \\
 *(f{\rm d}t + \mathbf a \cdot {\rm d}\mathbf r)
 = {\rm d}t \wedge \mathbf a \cdot {\rm d}\mathbf S + f{\rm d}V, \\
 *({\rm d}t \wedge \mathbf a \cdot {\rm d}\mathbf r + \mathbf b \cdot {\rm d}\mathbf S)
 = {\rm d}t \wedge \mathbf b \cdot {\rm d}\mathbf r - \mathbf a \cdot {\rm d}\mathbf S, \\
 *({\rm d}t \wedge \mathbf a \cdot {\rm d}\mathbf S + f{\rm d}V)
 = f{\rm d}t + \mathbf a \cdot {\rm d}\mathbf r, \\
 *(f{\rm d}t \wedge {\rm d}V)
 = -f.
\end{gather*}
\emph{Galilei} Hodge star:
\begin{gather} 
 \label{*GalOmega0}
 *f
 = f{\rm d}t \wedge {\rm d}V, \\
 \label{*GalOmega1}
 *(f{\rm d}t + \mathbf a \cdot {\rm d}\mathbf r)
 = {\rm d}t \wedge \mathbf a \cdot {\rm d}\mathbf S, \\
 \label{*GalOmega2}
 *({\rm d}t \wedge \mathbf a \cdot {\rm d}\mathbf r + \mathbf b \cdot {\rm d}\mathbf S)
 = {\rm d}t \wedge \mathbf b \cdot {\rm d}\mathbf r, \\
 \label{*GalOmega3}
 *({\rm d}t \wedge \mathbf a \cdot {\rm d}\mathbf S + f{\rm d}V)
 = f{\rm d}t, \\
 \label{*GalOmega4}
 *(f{\rm d}t \wedge {\rm d}V)
 = -f.
\end{gather}
\emph{Carroll} Hodge star:
\begin{gather} 
 \label{*CarOmega0}
 *f
 = f{\rm d}t \wedge {\rm d}V, \\
 \label{*CarOmega1}
 *(f{\rm d}t + \mathbf a \cdot {\rm d}\mathbf r)
 = f{\rm d}V, \\
 \label{*CarOmega2}
 *({\rm d}t \wedge \mathbf a \cdot {\rm d}\mathbf r + \mathbf b \cdot {\rm d}\mathbf S)
 = - \mathbf a \cdot {\rm d}\mathbf S, \\
 \label{*CarOmega3}
 *({\rm d}t \wedge \mathbf a \cdot {\rm d}\mathbf S + f{\rm d}V)
 = \mathbf a \cdot {\rm d}\mathbf r, \\
 \label{*CarOmega4}
 *(f{\rm d}t \wedge vV)
 = -f.
\end{gather}
In fact, for both new cases, Galilei as well as Carroll, there are as many as \emph{two}
 formulas for each degree of forms and some of them lead to zero operator.
 What is displayed here is always the \emph{non-zero} choice. See also Appendix~\ref{spacetimeHodge}.

\section{Galilei and Carroll vector spaces}
\label{galileicarrollvector}

\subsection{Galilei vector space -- basic facts}
\label{galileivectorbasics}

It is a triple $(V, \xi, h)$, where
\begin{enumerate}\itemsep=0pt
\item[--]$V$ is an $(n+1)$-dimensional vector space,
\item[--]$\xi$ is a non-zero covector (i.e., a $\binom01$-tensor) in $V$,
\item[--]$h$ is a rank-$n$ symmetric type-$\binom20$-tensor in $V$,
\item[--]such that $h (\xi, \ \cdot \ ) =0$.
\end{enumerate}
We call a frame $E_a = (E_0,E_i)$, $i=1,\dots ,n$
in $(V, \xi, h)$ and a (dual) coframe $E^a = \big(E^0,E^i\big)$ \emph{adapted} (or distinguished) if
\begin{enumerate}\itemsep=0pt
\item[--]$E^0 = \xi$,
\item[--]$h = \delta^{ij}E_i \otimes E_j$,
\end{enumerate}
so that, in (any) adapted frame, the components of the two tensors have ``canonical form''
\[ 
 \xi_a \leftrightarrow \left( \begin{matrix}
 \xi_0 \\
 \xi_i
 \end{matrix}
 \right)
 =
 \left( \begin{matrix}
 1 \\
 0
 \end{matrix}
 \right),
\qquad
 h^{ab} \leftrightarrow \left( \begin{matrix}
 h^{00} & h^{0i} \\
 h^{i0} & h^{ij}
 \end{matrix}
 \right)
 =
 \left( \begin{matrix}
 0 & 0 \\
 0 & \delta^{ij}
 \end{matrix}
 \right)
 =
 \left( \begin{matrix}
 0 & 0 \\
 0 & \mathbb I
 \end{matrix}
 \right).
\] 
The change-of-basis matrix $A$ between any pair $\hat E_a$, $E_a$ of adapted frames, given by $\hat E_a = A_a^bE_b$,
has the structure{\samepage
\begin{gather} \label{Galileichange-of-basismatrix}
 A^b_a \leftrightarrow \left( \begin{matrix}
 A^0_0 & A^0_i \\
 A^i_0 & A^i_j
 \end{matrix}
 \right)
 =
 \left( \begin{matrix}
 1 & 0 \\
 v^i & R^i_j
 \end{matrix}
 \right),
\qquad \text{i.e.,} \quad
 A \leftrightarrow \left( \begin{matrix}
 1 & 0 \\
 v & R
 \end{matrix}
 \right),
\end{gather}
where $R$ is $n$-dimensional rotation matrix.}

Such matrices form a Lie group $G$, subgroup of ${\rm GL}(n+1,\mathbb R)$,
the (homogeneous) \emph{Galilei group} ($R$ parametrizes rotations and $v$ Galilei boosts, respectively).

\subsection{Carroll vector space -- basic facts}\label{carrollvectorbasics}

It is a triple $\big(\tilde V, \tilde \xi, \tilde h\big)$, where
\begin{enumerate}\itemsep=0pt
\item[--] $\tilde V$ is an $(n+1)$-dimensional vector space,
\item[--]$\tilde \xi$ is a non-zero vector (i.e., a $\binom10$-tensor) in $\tilde V$,
\item[--]$\tilde h$ is a rank-$n$ symmetric type-$\binom02$-tensor in $\tilde V$,
\item[--]such that $\tilde h \big(\tilde \xi, \ \cdot \ \big) =0$.
\end{enumerate}
We call a frame $E_a = (E_0,E_i)$ and a (dual) coframe $E^a = \big(E^0,E^i\big)$, $i=1,\dots ,n$
in $\big(\tilde V, \tilde \xi, \tilde h\big)$ \emph{adapted} (or distinguished) if
\begin{enumerate}\itemsep=0pt
\item[--]$E_0 = \tilde \xi$,
\item[--]$\tilde h = \delta_{ij}E^i \otimes E^j$,
\end{enumerate}
so that, in (any) adapted frame, the components of the two tensors have ``canonical form''
\[ 
 \tilde \xi^a \leftrightarrow \left( \begin{matrix}
 \xi^0 \\
 \xi^i
 \end{matrix}
 \right)
 =
 \left( \begin{matrix}
 1 \\
 0
 \end{matrix}
 \right),
\qquad
 \tilde h_{ab} \leftrightarrow \left( \begin{matrix}
 \tilde h_{00} & \tilde h_{0i} \\
 \tilde h_{i0} & \tilde h_{ij}
 \end{matrix}
 \right)
 =
 \left( \begin{matrix}
 0 & 0 \\
 0 & \delta_{ij}
 \end{matrix}
 \right)
 =
 \left( \begin{matrix}
 0 & 0 \\
 0 & \mathbb I
 \end{matrix}
 \right).
\] 
The change-of-basis matrix $A$ between any pair $\hat E_a$, $E_a$ of adapted frames, given by $\hat E_a = A_a^bE_b$,
has the structure
\begin{gather} \label{Carrollchange-of-basismatrix}
 A^b_a \leftrightarrow \left( \begin{matrix}
 A^0_0 & A^0_i \\
 A^i_0 & A^i_j
 \end{matrix}
 \right)
 =
 \left( \begin{matrix}
 1 & v_i \\
 0 & R^i_j
 \end{matrix}
 \right),
\qquad \text{i.e.,} \quad
 A \leftrightarrow \left( \begin{matrix}
 1 & v^{\rm T} \\
 0 & R
 \end{matrix}
 \right),
\end{gather}
where $R$ is $n$-dimensional rotation matrix.

Such matrices form a Lie group $G$, subgroup of ${\rm GL}(n+1,\mathbb R)$,
the (homogeneous) \emph{Carroll group} ($R$ parametrizes rotations and $v$ Carroll boosts, respectively).

It follows that \emph{boosts} formulas $x'^a = A^a_bx^b$ lead to physically strange looking
(yet well-known) expressions (\ref{boostcarinft}) and (\ref{boostcarinfr}).

\subsection{Forms on Galilei/Carroll vector space}\label{formsgalileivector}

In Appendices~\ref{formsgalileivector} and \ref{intertwinformsgalileivector}
common treatment of forms on both vector spaces is given.
The group~$G$ is either Galilei group from Appendix \ref{galileivectorbasics}
or Carroll group from Appendix~\ref{carrollvectorbasics}.
So whenever matrix~$A$ is present in a formula,
we mean \emph{either} (in Galilei case) the one from (\ref{Galileichange-of-basismatrix})
\emph{or} (in Carroll case) the one from (\ref{Carrollchange-of-basismatrix}).

There is a natural right (free) action of the group $G$ on the set $\mathcal E$ of adapted frames
\[ 
 R_A\colon\ \mathcal E \to \mathcal E,
\qquad
 (R_AE)_a := A^b_aE_b
\] 
and, consequently, on coframes $\mathcal E^*$ \big($E^a \mapsto \big(A^{-1}\big)^a_bE^b$\big).
This induces representation $\rho_p$ of $G$ on the space of $p$-forms on $V$,
\begin{gather*}
 \rho_p\colon \ G \to \Aut \Lambda^pV^*,
\\
 \rho_p(A)\left(
 \frac{1}{p!}\alpha_{a\dots b}E^a \wedge \dots \wedge E^b
 \right)
 :=
 \frac{1}{p!}\alpha_{a\dots b}\big(A^{-1}\big)^a_cE^c \wedge \dots \wedge \big(A^{-1}\big)^b_dE^d,
\end{gather*}
and, consequently, on the (isomorphic) space $\hat \Lambda^p$ of \emph{components} of $p$-forms on $V$
\begin{gather} \label{Greproncomppforms1}
 \hat \rho_p\colon\ G \to \Aut \hat \Lambda^p,
\\ 
 \hat \rho_p(A)\colon \
 \alpha_{a\dots b} \
 \mapsto \
 \big(A^{-1}\big)^c_a \cdots \big(A^{-1}\big)^d_b \alpha_{c\dots d}.\nonumber
\end{gather} 
It is well known (see, e.g.,\ \cite{Feckoangl, Sternbergdifgeom, Trautman})
that the space $\Lambda^pV^*$ of $p$-forms in $V$ is canonically isomorphic to the space of \emph{equivariant} maps
\begin{gather} \label{Gequivcomppforms}
 \alpha_p \leftrightarrow \hat \alpha_p,
\qquad
 \hat \alpha_p \colon \
 \mathcal E \to \hat \Lambda^p,
\qquad
 \hat \alpha_p \circ R_A = \hat \rho_p\big(A^{-1}\big) \circ \hat \alpha_p.
\end{gather}

\subsection{Intertwining operators on forms on Galilei/Carroll vector space}
\label{intertwinformsgalileivector}

Let $\hat \alpha_p$ be an equivariant map (\ref{Gequivcomppforms})
corresponding to a $p$-form $\alpha_p$ in $V$ and let
\begin{gather} \label{hataqp}
 \hat a_{qp} \colon \ \hat \Lambda^p \to \hat \Lambda^q
\end{gather}
be an \emph{intertwining} operator (equivariant linear mapping) for representations $\hat \rho_p$ and $\hat \rho_q$,
respectively, i.e., obeying
\begin{gather} \label{hataqpobeys}
 \hat \rho_q (A) \circ \hat a_{qp} = \hat a_{qp} \circ \hat \rho_p (A).
\end{gather}
Then a simple check shows that mere \emph{composition} of maps
\[ 
 \mathcal E \overset{\hat \alpha_p} \to \hat \Lambda^p \overset{\hat a_{qp}} \to \hat \Lambda^q
\] 
produces an equivariant map corresponding to a $q$-form in $V$.

So we can reconstruct corresponding operators \emph{on forms} (transforming $p$-forms to $q$-forms)
\begin{gather} \label{alphaptoqforms}
 \hat \alpha_p \mapsto \hat \beta_q := \hat a_{qp} \circ \hat \alpha_p
\end{gather}
once we know intertwining operators $\hat a_{qp}$ between the representations $\hat \rho_p$ and $\hat \rho_q$.

\section{Galilean and Carrollian manifolds}
\label{galileancarrollianmanifold}

\subsection{Galilean/Carrollian structure}
\label{galileancarrollianstructure}

Here, again, the formal procedure is the same for both Galilean an Carrollian cases.
So when we write, say, ``Galilean'', one can replace it by ``Carrollian'' as well
(simply imagine matrix $A$ being either from the former or from the latter group).

On \emph{Galilean manifold} $(M, \xi,h)$, each \emph{tangent space} $T_mM$ is endowed
with the structure of \emph{Galilei} vector space described in Appendix~\ref{galileivectorbasics}.

So the \emph{Galilean structure} on $M$ may be (standardly) described as a $G$\emph{-structure}
(see, e.g., \cite{Feckoangl, Kunzle1972, Sternbergdifgeom, Trautman}),
i.e., in terms of a principal $G$-bundle
\[
 \pi \colon\ P\to M.
\]
Here points in the fiber over $m\in M$ refer to \emph{adapted} frames in $T_mM$.
So the principal bundle may be regarded as a \emph{restriction} of the frame bundle $LM$
to the sub-bundle of adapted frames.

There is a natural right (free, vertical and in each fibre transitive) action of the group~$G$
(homogeneous Galilei group) on $P$
\[
 R_A\colon \ P \to P,
\qquad
 \pi \circ R_A = \pi.
\]

\subsection{Forms on Galilean/Carrollian manifold}\label{formsongalileanmanifold}

In general, \emph{geometric quantities} on $M$ may be described either as sections of \emph{associated} bundles or,
equivalently (see, e.g., \cite{Feckoangl, Trautman}), as \emph{equivariant} functions on the \emph{principal} bundle $P$,
i.e., as mappings
\[ 
 \Phi \colon \ P \to (W,\hat \rho),
\qquad
 \Phi (eA) = \hat \rho \big(A^{-1}\big) \Phi (e),
\qquad \text{i.e.,}\quad
 R_A^*\Phi = \hat \rho \big(A^{-1}\big) \circ \Phi.
\] 
Here $(W,\hat \rho)$ is a representation module of $G$.
The corresponding quantity (described by $\Phi$) is called
 \emph{quantity} (on $M$) \emph{of type} $\hat \rho$.\footnote{$\Phi (e)\in W$ plays the role of ``components'' of the quantity under consideration
 w.r.t.\ the frame $e\in P$.
 The ``actual'' (frame-dependent) quantity $\varphi$ living on $M$ is computed from $\Phi$
 by pull-back $\varphi = \sigma^*\Phi$ w.r.t.\ a local section~$\sigma$,
 i.e., a choice of local frame field on~$M$.}

In particular, for representation $\hat \rho_p$ in $\hat \Lambda^p$ from~(\ref{Greproncomppforms1}),
the corresponding quantities of type $\hat \rho_p$ are nothing but $p$\emph{-forms} on the Galilean manifold $(M, \xi,h)$.
So, we can treat elements of $\Omega^p(M)$ as mappings
\begin{gather} \label{functionsoftyperho_ponP}
 \Phi_p \colon \ P \to \big(\hat \Lambda^p,\hat \rho_p\big),
\qquad
 \Phi_p (eA) = \hat \rho_p \big(A^{-1}\big) \Phi_p (e).
\end{gather}

\subsection{Intertwining operators on forms on Galilean/Carrollian manifold}
\label{intertwinonformsonmanifold}

Let $\Phi_p$ be an equivariant map (\ref{functionsoftyperho_ponP})
corresponding to a $p$-form $\alpha_p$ on $M$ and let $\hat a_{qp}$
be an operator from (\ref{hataqp}).

Then a simple check shows that mere \emph{composition} of maps
\[
 P \overset{\Phi_p} \to \hat \Lambda^p \overset{\hat a_{qp}} \to \hat \Lambda^q
\]
produces an equivariant map corresponding to a $q$-form on $M$.

So we can reconstruct corresponding operators on \emph{forms} on a \emph{Galilean manifold}
(mappings $\Omega^p(M) \to \Omega^q(M)$, transforming $p$-forms to $q$-forms on $M$){\samepage
\begin{equation} \label{alphaptoqformsmanifold}
 \Phi_p \mapsto \hat a_{qp} \circ \Phi_p
\end{equation}
once we know matrices $\hat a_{qp}$, intertwining operators between the representations $\hat \rho_p$ and $\hat \rho_q$.}

In this way, a simple result from linear algebra (of Galilei \emph{vector space}) eventually becomes
a~much more interesting result concerning operators acting on \emph{differential} forms on \emph{any} \emph{Galilean manifold} $(M,\xi,h)$).\footnote{The composition in (\ref{alphaptoqformsmanifold}) simply means
 that \emph{components} (w.r.t.\ \emph{any adapted} local frame) of differential forms
 are scrambled by (constant) matrices $\hat a_{qp}$.}

\section{Galilei/Carroll spacetime versus vector space computations}
\label{spacetimevectorspace}

Each (finite dimensional) vector space $V$ may be regarded as a manifold
(components $x^a$ of vectors w.r.t.\ a frame $e_a$ may be used as global coordinates, $v\leftrightarrow x^a$).

The linear structure of the space enables one to identify each tangent space $T_vV$ with the space $V$ itself.
(Canonical isomorphism $V\to T_vV$ reads $w\mapsto \dot \gamma (0)$ for $\gamma (t)= v+wt$.)

Consequently, whenever we have a type $\binom pq$ tensor $B$ in $V$,
we can canonically associate with it a type $\binom pq$ tensor \emph{field} $b$ on $V$ (as a manifold).
The correspondence reads
\begin{equation} \label{correspondenceVV}
 e^a \leftrightarrow {\rm d}x^a,
\qquad
 e_a \leftrightarrow \partial_a,
\end{equation}
i.e.,{\samepage
\[ 
B \equiv B_{a\dots }^{\ \dots b} e^a \otimes \dots \otimes e_b
 \leftrightarrow
 B_{a\dots }^{\ \dots b} {\rm d}x^a\otimes \dots \otimes \partial_b \equiv b
\] 
(components $B_{a\dots }^{\ \dots b}$ are the same in both expressions and \emph{constant}).}

Now if a linear mapping $A\colon V\to V$
preserves (say) a type $\binom 02$ tensor $B$ (bilinear form) in the usual sense of linear algebra,
\begin{gather} \label{ApreservesBlinalg}
 B(Au,Aw)=B(u,w),
\qquad \text{i.e.,} \quad
 A_a^cB_{cd}A^d_b=B_{ab}
\end{gather}
then, on $V$ treated as a manifold, we speak of a smooth mapping
\[
 f\equiv f_A\colon \ V\to V
\qquad \text{in coordinates} \quad
 x^a\mapsto A^a_bx^b
\]
and the tensor field $b$ is preserved in the sense of \emph{pull-back}
\begin{gather} \label{ApreservesBmanifold}
 f_A^* b=b,
\end{gather}
i.e., as we usually understand the concept of ``preserving a geometrical quantity'' on manifolds.

Preserving of $B$ may also be treated \emph{infinitesimally}.
In linear algebra, we get from (\ref{ApreservesBlinalg}), for operators of the form $A=\mathbb I +\epsilon C$,
\[ 
 B(Cu,w)+B(u,Cw)=0,
\qquad \text{i.e.,}\quad
 C_a^cB_{cb} + B_{ac}C^c_b =0.
\] 
In the language of manifolds, we get from (\ref{ApreservesBmanifold}) in a standard way
\[ 
 \mathcal L_\xi b=0,
\qquad
 \xi = x^bC_b^a\partial_a,
\] 
where $L_\xi$ is the \emph{Lie derivative}.
The specific structure of $\xi$ (\emph{linear} vector field, matrix $C$ is constant)
is dictated by the fact that the corresponding flow is to consist of linear transformations.

So, summing up, the computations of the type mentioned above may be either realized
in the language of linear algebra or in the language of (``linear'') differential geometry.

Now notice that our computations in Sections \ref{generrot} and \ref{generboost},
leading to explicit expressions (\ref{lorgalcar0comp})--(\ref{lorgalcar4comp})
for representations generators $\rho_p$ listed in Section \ref{generrotboost},
are actually performed just within the above mentioned ``linear differential geometry'':
Formulas like (\ref{rotdt})--(\ref{rotdr}) for rotations or
(\ref{lorgalcarboostinfdt})--(\ref{lorgalcarboostinfdr}) for boosts
become exactly (infinitesimal versions of) $\hat E^a = \big(A^{-1}\big)^a_bE^b$ (with $A = \mathbb I + \epsilon C$)
in the language of Appendix~\ref{galileivectorbasics}
\emph{once they are translated} (via vocabulary~(\ref{correspondenceVV})) into the language
of the Galilei \emph{vector space}.
\big(Standard coordinate coframe ${\rm d}x^a \equiv ({\rm d}t,{\rm d}x^i)$ becomes an \emph{adapted} coframe $E^a \equiv \big(E^0,E^i\big)$.\big)

Put it differently, the operators listed in Section \ref{generrotboost}
are exactly those operators mentioned in (\ref{Greproncomppforms1}) for which
intertwining operators $\hat a_{qp}$ are computed via solving (\ref{hataqpobeys}).
More precisely, they are generators of the \emph{derived} representation $\hat \rho'_q$
of the Lie algebra $\mathcal G$ of the group $G$, given standardly by
\[ 
 \hat \rho_q (\mathbb I + \epsilon C) \doteq \hat 1 + \epsilon \hat \rho'_q(C).
\] 
Since, however, solutions $\hat a_{qp}$ of (\ref{hataqpobeys}) do not change
when $\hat \rho_q (A)$ is replaced by $\hat \rho'_q(C)$, our intertwining operators
listed (in the form of \emph{matrices}) in Sections \ref{intertwinminknontrivial}--\ref{intertwincarnontrivial}
are exactly those mentioned in equation (\ref{hataqpobeys}).

Then, however, (\ref{alphaptoqforms}) describes the corresponding operators on forms
(still in the Galilei vector space) and finally (\ref{alphaptoqformsmanifold})
describes the operators on \emph{forms} on \emph{Galilean manifold}.

\section[Explicit expressions of *-operator on Galilean/Carrollian spacetimes]{Explicit expressions of $\boldsymbol{*}$-operator on Galilean/Carrollian\\ spacetimes}
\label{spacetimeHodge}

Due to (\ref{alphaptoqformsmanifold}), our Hodge star operator,
when acting on forms on \emph{any} Gali\-le\-an/Car\-rollian spacetime,
scrambles components of forms w.r.t.\ \emph{any adapted} frame by \emph{the same} matrices,
namely by~$\hat a_{qp}$. Since we know how these matrices look like
for forms on special Galilean/Carrollian spacetime, the Galilei/Carroll spacetime
(see (\ref{*GalOmega0})--\ref{*CarOmega4})), we are simply to replace
our particular adapted (global, holonomic) coframe field ${\rm d}x^a \equiv \big({\rm d}t, {\rm d}x^i\big)$ from Galilei/Carroll spacetime
with a \emph{general adapted} (possibly local and non-holonomic) \emph{coframe field} $e^a \equiv \big(e^0, e^i\big)$
living on the Galilean/Carrollian spacetime under consideration.
Recall that they are defined (see Appendices \ref{galileivectorbasics} and \ref{carrollvectorbasics}),
for a Galilean manifold $(M, \xi, h)$ and a~Carrollian manifold $\big(M, \tilde \xi, \tilde h\big)$,
by the properties
\begin{alignat*}{3} 
 &\text{Galilean frame/coframe field}\colon \ && e^0 = \xi, \qquad h = \delta^{ij}e_i \otimes e_j,& \\
 &\text{Carrollian frame/coframe field}\colon \ &&e_0 = \tilde \xi, \qquad \tilde h = \delta_{ij}e^i \otimes e^j.&
\end{alignat*}
Then the above mentioned formulas (\ref{*GalOmega0})--\ref{*CarOmega4}) with \emph{substitutions}
\begin{gather*} 
 {\rm d}t
 \mapsto e^0, \\
 \mathbf a \cdot {\rm d}\mathbf r
 \mapsto a_1e^1 + a_2e^2 + a_3e^3, \\
 \mathbf a \cdot {\rm d}\mathbf S
 \mapsto a_1{\rm d}S_1 + a_2{\rm d}S_2 + a_3{\rm d}S_3
 \equiv a_1e^2\wedge e^3 + a_2e^3 \wedge e^2 + a_3e^1 \wedge e^3, \\
{\rm d}V
 \mapsto e^1 \wedge e^2 \wedge e^3
\end{gather*}
provide valid expressions of our Hodge star operator on forms on \emph{any} Galilean/Carrollian spacetime.

And just for the sake of completeness, let's state how the degree $\pm 1$ operators look like:
\[
 \xi \wedge (\dots) \quad \text{on} \ \Omega \ (\text{Galilei}),
\qquad
 i_{\tilde \xi} (\dots) \quad \text{on} \ \Omega \ (\text{Carroll}).
\]

\subsection*{Acknowledgements}

The author acknowledges support from grant VEGA 1/0703/20 and sincerely
thanks the anonymous referees whose remarks contributed to the quality of the paper.

\pdfbookmark[1]{References}{ref}
\LastPageEnding

\end{document}